\documentclass[letterpaper]{JHEP3}

\usepackage{epsfig}

\def\Box{{\hbox{$\sqcup$}\llap{\hbox{$\sqcap$}}}}

\def \lsim{\mathrel{\vcenter
     {\hbox{$<$}\nointerlineskip\hbox{$\sim$}}}}
\def \gsim{\mathrel{\vcenter
     {\hbox{$>$}\nointerlineskip\hbox{$\sim$}}}}

\def\be{\begin{equation}}
\def\ee{\end{equation}}
\def\bea{\begin{eqnarray}}
\def\eea{\end{eqnarray}}
\def\nn{\nonumber}

\def\pref#1{(\ref{#1})}
\def\eqref#1{(\ref{#1})}
\def\endignore{}
\def\ignore #1\endignore{} 

\def\bd{\begin{displaymath}}
\def\ed{\end{diplaymath}}

\def\eff{{\rm eff}}

\def\ie{{\it i.e.\ }}

\def\cA{{\cal A}}

\def\cD{{\cal D}}
\def\cH{{\cal H}}
\def\cL{{\cal L}}

\def\ls{\lambda_{2\star}}
\def\lsa{|\lambda_{2\star}|}

\def\ba{\begin{eqnarray}}
\def\ea{\end{eqnarray}}
\def\be{\begin{equation}}
\def\ee{\end{equation}}

\def\ssB{{\scriptscriptstyle B}}
\def\ssH{{\scriptscriptstyle H}}
\def\ssK{{\scriptscriptstyle K}}
\def\ssL{{\scriptscriptstyle L}}
\def\ssM{{\scriptscriptstyle M}}

\def\ssT{{\scriptscriptstyle T}}
\def\ssW{{\scriptscriptstyle W}}
\def\ssY{{\scriptscriptstyle Y}}
\def\ssZ{{\scriptscriptstyle Z}}

\def\o{\omega}

\def\O{\mathcal{O}}

\def\nn{\nonumber}

\def\d{\mathrm{d}}

\def\e{\epsilon}

\def\({\left(}
\def\){\right)}
\def\ie{{\it i.e. }}

\def\bd{\boxdot}
\def\text#1{\hbox{\rm {#1}}}

\def\pd{\partial}
\def\lt{\bar \lambda_2}
\def\lf{\bar \lambda_4}
\def\ie{{\it i.e. }}

\title{The Hierarchy Problem and the Self-Localized Higgs}

\author{C.P. Burgess,${}^{1,2,3}$ Claudia de Rham${}^{1,2}$
and Leo van Nierop${}^2$ \\
${}^1$ Perimeter Institute for Theoretical Physics, Waterloo ON,
N2L 2Y5, Canada.\\
${}^2$
Physics \& Astronomy, McMaster University, Hamilton ON, L8S 4M1,
Canada. \\
${}^3$ Theory Division, CERN, CH-1211 Geneva 23, Switzerland. \\
}

\date{}

\abstract { We examine brane-world scenarios in which all the
observed Standard Model particles reside on a brane but the Higgs
is an elementary extra-dimensional scalar in the bulk. We show
that, for codimension 2 branes, often-neglected interactions
between the bulk Higgs and the branes cause two novel effects.
First, they cause $\langle H \rangle$ to depend only
logarithmically on the UV-sensitive coefficient, $m_\ssB^2$, of
the mass term, $m_\ssB^2 \, H^*H$, of the bulk potential, thus
providing a new mechanism for tackling the hierarchy problem.
Second, the Higgs brane couplings cause the lowest mass KK mode to
localize near the brane without any need for geometrical effects
like warping. We explore some preliminary implications such models
have for the Higgs signature at the LHC, both in the case where
the extra dimensions arise at the TeV scale, and in ADD models
having Large Extra Dimensions. Novel Higgs features include
couplings to fermions which can be different from Standard Model
values, $m_f/v$, despite the fermions acquiring their mass
completely from the Higgs expectation value. }

\begin{document}

\section{Introduction}

In the Standard Model (SM) the Higgs field is in many ways the odd
man out. In the absence of the Higgs the only interactions that
remain are gauge interactions, characterized by only a handful of
coupling constants. But with the Higgs comes the deluge of
parameters that parameterize our ignorance of the ultimate origins
of the model's many masses and mixing angles. And among these
parameters is the one dimensionful quantity, $\mu$, that governs
the size of the $\mu^2 H^*H$ term in the Higgs potential, and by
fixing the size of the Higgs v.e.v. sets the scale for all masses.
It is the sensitivity of this parameter to much heavier scales
that is at the root of the hierarchy problem \cite{SM}.

Historically, the hierarchy problem has been one of the main
motivations for exploring brane-world scenarios for physics beyond
the Standard Model \cite{ADD,RS}, for which all of the observed SM
particles are trapped on a (3+1)-dimensional brane within an
extra-dimensional bulk. Motivated by the observations that the
Higgs is the lone SM particle yet to be observed, we here explore
the idea that it is the only SM particle that is {\it not}
confined to a brane: {\it i.e.} whereas all other SM particles
live on a brane, the Higgs lives in the bulk. The hope is that
this might account for its special role within the SM.

Brane-world models with the Higgs in the bulk have been examined
in the literature, most often within the context of 5D
Randall-Sundrum constructions \cite{RS}. Yet these models differ
from the present proposal in one of two ways: either by imagining
the extra-dimensional Higgs to be related to other fields by
supersymmetry \cite{Extra5DHiggs,Extra6DHiggs}; or by taking the
Higgs to be the 4D scalar component of what is `really' an
extra-dimensional gauge potential
\cite{Gauge5DHiggs,Gauge6DHiggs}. The motivation for doing so is
the expectation that the extra-dimensional gauge symmetries can
help alleviate the hierarchy problem, potentially allowing some of
the properties of Higgs interactions to be unified with those of
the gauge interactions. Implicit in this is the belief that a
Higgs that is a bona-fide extra-dimensional scalar makes no
progress towards alleviating the hierarchy problems of the usual
4D Higgs.

A model more similar to the one studied here was considered in
ref.~\cite{Dudas1,Dudas2}, although from a different point of
view. In ref.~\cite{Dudas1} the authors study the effects of
codimension-2 brane couplings on a massless bulk scalar, with a
focus on couplings close to the critical value for which the
symmetry-breaking properties of the vacuum change.
Ref.~\cite{Dudas2} generalizes to massive bulk fields, but without
the focus of this paper on the hierarchy problem, and consequently
without the study of couplings to fermions and gauge bosons
described herein.

It is simple to see why extra dimensions in themselves are
generally believed not to alleviate the hierarchy problem. This is
because the Higgs potential,
\be
    U = - \frac{m_\ssB^2}{2} \, H^*H + \frac{g}{4} \, (H^* H)^2 \,,
\ee
\pagebreak
is always minimized by $H^* H = m_\ssB^2/g$, where in $n$
dimensions $g$ has the (engineering) dimension of (mass)${}^{4-n}$
while $m_\ssB$ always simply has the dimension of mass. But the
essence of the hierarchy problem is that because $m_\ssB$ is
proportional to a positive power of mass, it generically receives
contributions from heavy particles that grow with the mass, $M$,
of the particles involved, and so is dominantly affected by the
heaviest such particle that can contribute. Since $m_\ssB$ is a
positive power of mass in any number of dimensions it is hard to
see how the hierarchy problem can be ameliorated simply by placing
the Higgs into the bulk.

In this paper we show why this simple argument is incorrect once
the couplings between a bulk Higgs and the brane are properly
taken into account. The brane-bulk interactions change the
argument because the Higgs potential on the brane, $U_b$, and in
the bulk, $U_\ssB$, can disagree on which value of the Higgs
v.e.v. has the least energy. In this case the system generically
resolves this potential frustration by appropriately balancing
these potential energies with the gradient energies which punish
the field for attempting to interpolate between the two minima.
But {\it if} the brane has codimension 2 ({\it i.e.} there are two
dimensions transverse to the brane, such as for a
(3+1)-dimensional brane situated in a 6D bulk), the Higgs likes to
vary logarithmically near the branes, and the gradient energy
associated with this variation is such that the resulting v.e.v.
only depends {\it logarithmically} on the UV-sensitive term,
$m_\ssB$, of the bulk potential. Braneworld models can help with
naturalness problems for a number of reasons; brane-bulk couplings
provide a new way for them to do so.
We show that the lunch is nevertheless not completely free,
however, since the hierarchy problem gets partially recast as a
requirement for the coefficients of the brane interactions $(H^*
H)^2$ and $D_\ssM H^* D^\ssM H$ being required to be suppressed by
very different scales. This kind of hierarchical suppression
usually does not arise between two operators like these, that are
not distinguished by low-energy symmetries or selection rules.

We also show how Higgs-brane interactions change another
fundamental piece of widely-held intuition regarding the
properties of a bulk Higgs. In the presence of a (positive)
extra-dimensional mass term, $U_\ssB = +\frac12 \, m_\ssB^2 H^*H$,
the spectrum of Kaluza-Klein (KK) states would usually be expected
to consist of a multitude of levels (generically spaced by $M_c
\sim 2\pi/L$ for a toroidal extra dimension of circumference $L$,
say) that start at energies above a gap, $m_k \ge m_\ssB$. We show
here that brane-Higgs interactions can generically introduce a
state which lives within this gap, $m < m_\ssB$, that is `bound'
in the sense that its wave-function is localized at the position
of the brane. We call this the `self-localized' state inasmuch as
its localization is a consequence only of the Higgs
self-interactions and not on any geometric effects, such as those
due to warping.

These arguments are presented in more detail in their simplest
context in the next section, $\S2$. $\S3$ then tries to fashion an
approach to the hierarchy problem by providing a preliminary
discussion of the kinds of interactions that would be required for
a realistic model, and the ways in which the low-energy Higgs
couplings resemble and differ from those of the SM Higgs, as a
function of the scales involved. $\S4$ then follows with a
discussion of some of the potential signatures and constraints
such a scenario might have for Higgs physics. Our conclusions are
briefly summarized in $\S5$.

\section{Vacuum Energetics of Extra-Dimensional Scalars}

In this section we describe the interplay between brane and bulk
energetics for the simplest toy model: a single real scalar,
$\phi$, in the presence of both brane and bulk potentials, $U_b$
and $U_\ssB$. We first review the more familiar situation of a
codimension-1 brane in a 5D bulk, and then contrast this with the
codimension-2 case with 6 bulk dimensions. (The situation for
higher codimension is sketched in Appendix \ref{AppHigherCod}.)
Because they are peripheral to our main point we neglect
gravitational effects in what follows, and so assume the mass
scales involved are low enough for this to represent a good
approximation.

\subsection{Codimension-one}

We first consider the codimension-one case, reproducing the
results of ref.~\cite{cod1case}. Consider the following 5D scalar
field theory, having both bulk- and brane-localized interactions,
\ba
    S = -\int \d^4 x\, \d y \, \left[ \frac12 (\partial_\ssM \phi
    \, \partial^\ssM \phi ) + U_\ssB
    (\phi)+\delta(y)\, U_b(\phi) \right] \,,
\ea
with $\{ x^\ssM \} = \{ x^\mu, y \}$. The field equation for this
model is
\be
    \partial^\ssM \partial_\ssM \phi - U_\ssB'(\phi)
    = \delta(y)U_b'(\phi) \,,
\ee
and the integration of this equation across the brane position
(assuming continuity of $\phi$) further implies the scalar jump
condition
\be \label{cod1jumpcondition}
    \left[ \partial_y \phi \right]_0 = U_b'(\phi_0) \,,
\ee
where $\phi_0 = \phi(y=0)$ and $[A]_0 = A(y=0^+) - A(y=0^-)$. The
classical energy density per unit brane volume associated with a
given field configuration in this model is then
\ba
    {\mathcal H}=\int_{y_{\rm min}}^{y_{\rm max}}
    \d y \left[ \frac12 \left( \dot\phi^2
    + (\nabla \phi)^2 + (\pd_y \phi)^2 \right) +U_\ssB(\phi) \right] +
    U_b(\phi_0)\,,
\ea
where $y_{\rm min} < 0 < y_{\rm max}$ and $\nabla$ denotes
differentiation in the in-brane spatial directions, $\{x^i \}$.

We now specialize to the case where the field has only a mass term
in the bulk, while it has a quartic interaction on the brane.
Keeping in mind that $\phi$ has dimension (mass)${}^{3/2}$ in 5
dimensions,
\be
    U_\ssB(\phi) = \frac12 \, m_\ssB^2 \, \phi^2 \quad \text{and}
    \quad
    U_b(\phi) = -\frac12\,m_b \, \phi^2 + \frac{1}{4
    M_b^2} \, \phi^4\,,
\ee
where $m_b > 0$ is chosen to ensure that the minimum of the brane
potential occurs at the nonzero value $\phi^2 = M_b^2 m_b$, in
contrast with the bulk potential which is minimized at $\phi = 0$.

Since $U_b$ and $U_\ssB$ are not minimized by the same
configuration, the vacuum solution need not correspond to a
constant field configuration, $\partial_\ssM \phi = 0$. Since the
solutions to the field equations that only depend on $y$ are
exponentials, $\phi \propto e^{\pm m_\ssB y}$, the general bulk
solution is a linear combination of such terms. If the extra
dimension is sufficiently large --- $|m_\ssB y_{\rm min}| \gg 1$
and $m_\ssB y_{\rm max} \gg 1$ --- then we can drop the solutions
which grow exponentially far from the brane, just as if the extra
dimension were noncompact. In this case the vacuum configuration
should vanish at infinity, and the solution is therefore given by
\ba
    \phi(y)=\bar\phi \, e^{- m_\ssB |y|}\,,
\ea
where $\bar \phi$ is to be fixed using the boundary condition,
eq.~\pref{cod1jumpcondition}, at $y=0$: \ie $-2 m_\ssB \bar\phi =
U_b'(\bar\phi)$, or
\be
    \left( 2 m_\ssB - m_b + \frac{\bar\phi^2}{M_b^2} \right)
    \bar\phi = 0 \,.
\ee
When $m_b < 2 \, m_\ssB$ the only real solution allowed is
$\bar\phi = 0$, but when $m_b > 2\, m_\ssB$ there are three
solutions for $\bar\phi$, corresponding to $\bar\phi = 0$ and
$\bar\phi = \pm \phi_c$, with
\be
    \phi_c^2 = M_b^2 \( m_b - 2 m_\ssB \) \,.
\ee

Since $\cH = 0$ for $\bar\phi = 0$ and $\cH = -\frac14 M_b^2 \(m_b
- 2 m_\ssB\)^2$ for $\bar\phi^2 = \phi_c^2$, we see that it is the
nontrivial configuration which represents the classical ground
state when $m_b > 2\, m_\ssB$. This can also be seen more
generally by writing the energy density as a function of $\bar
\phi$,
\be
    \mathcal H (\bar \phi) = -\frac 12 \(m_b-2m_\ssB\) \bar \phi^2
    +\frac{1}{4 M_b^2} \, \bar \phi^4\,,
\ee
which is indeed minimized, for $m_b > 2m_\ssB$, by $\bar\phi = \pm
M_b \sqrt{m_b-2m_\ssB}$, with the unstable stationary point, $\bar
\phi = 0$, situated at a local maximum.

The resulting vacuum
\be
    \phi^2(y)= M_b^2 \(m_b-2m_\ssB\) e^{-2 m_\ssB |y|} \,,
\ee
extrapolates from the bulk minimum ($\phi = 0$) for large $y$ to
the value $\phi_0 = \pm M_b \sqrt{m_b - 2m_\ssB}$ at the brane.
This represents a compromise between the bulk minimum, the value
$\bar\phi = \pm M_b \sqrt{m_b}$, which minimizes $U_b$, and the
gradient energy required to interpolate between the two. Notice
that $\phi_0$ approaches the brane minimum in the limit where the
bulk potential is very flat, $m_\ssB \ll m_b$.

\subsection{Codimension-two}

We now contrast the previous results with a similar analysis for
the codimension-2 case of a real scalar field coupled to a 3-brane
in 6 spacetime dimensions, where we show that the larger gradient
energy more strongly favors the minimum of the bulk potential
relative to that of the brane. Using the action
\ba
    S = -\int \d^4 x\, \d^2y  \left[
    \frac12 (\partial_\ssM \phi \, \partial^\ssM \phi)
    + U_\ssB (\phi) + \delta^2(y) \, U_b(\phi) \right] \,,
\ea
we have the equation of motion
\be \label{cod2eom}
    \partial^\ssM \partial_\ssM \phi - U_\ssB'(\phi)
    = \delta^2(y) U_b'(\phi) \,.
\ee
Assuming a flat space-time metric
\ba
    \d s^2=\eta_{\mu\nu}\d x^\mu \d x^\nu+\d r^2+r^2\d \theta^2\,,
\ea
and integration of the equation of motion across a very small disc centered on
the brane position at $r=0$ (assuming continuity of $\phi$) further implies
the condition
\be \label{cod2jumpcondition}
    \lim_{r \to 0} \left[2\pi r \partial_r \phi \right]
    = U_b'(\phi_0) \,,
\ee
where $r$ measures the radial distance from the brane situated at
$r=0$. For configurations depending only on $r$, this corresponds
to using the radial field equation
\ba \label{eom}
    \frac1r \, \partial_r \Bigl( r\, \partial_r\phi \Bigr)
    -U_\ssB'(\phi) = \frac{\delta_+(r)}{2 \pi r} \, U_b'(\phi)\,,
\ea
where $\delta_+(r)$ is normalized so that $\int_0^a \d r
\delta_+(r) f(r) = f(0)$, for any $a > 0$.

Since our interest is in how the system resolves the frustration
of minimizing brane and bulk potentials having different minima,
we specialize to the simple choices
\be
    U_\ssB(\phi) = \frac12 \, m_\ssB^2 \phi^2 \quad \text{and}\quad
    U_b(\phi) = -\frac 12 \, \lambda_2 \, \phi^2
    +\frac14 \, \lambda_4 \, \phi^4\,,
\ee
with both $\lambda_2$ and $\lambda_4$ taken to be positive.
Keeping in mind a 6D scalar field has dimension (mass)${}^2$, we
see that the parameter $\lambda_2$ is dimensionless, while
$\lambda_4 = 1/M_b^4$.

Provided the extra dimensional radius, $L$, satisfies $m_\ssB L
\gg 1$, it is a good approximation to demand the bulk vacuum
configuration to vanish at large $r$, leading to the following
solution
\ba \label{bulkprofile}
    \phi(r) = \bar\phi \, K_0 (m_\ssB r) \,,
\ea
where the modified Bessel function, $K_0(z)$, falls exponentially
with $z$ for large $z$ and diverges logarithmically as $z$
approaches zero. Using $K_0(z) = - \ln (z/2) - \gamma + O(z)$ to
evaluate $r \partial_r \phi \to -  \bar\phi$ as $r \to 0$, allows
the boundary condition, eq.~\pref{cod2jumpcondition}, to be
written
\be
    -2\pi \bar\phi = U_b'(\phi_0) \,,
\ee
and here we encounter the first difference from the codimension-2
case: $\phi(r)$ diverges logarithmically as $r \to 0$, making
$\phi_0 = \phi(r=0)$ ill defined. Regularizing\footnote{This
regularization can be done more precisely by modelling the
codimension-2 brane by a small codimension-1 circle at radius
$r=\e$, and using the codimension-1 jump conditions to relate the
exterior bulk fields to the nonsingular fields in the circle's
interior \cite{JJ,UVCaps}. } by evaluating at a small but nonzero
radius, $r = \e$, gives $\phi_\e = \bar\phi \, z_\e$, where
\be
    z_\e \equiv K_0(m_\ssB \e) = \ell
    + \ln 2 - \gamma + \O(\e) \,,
\ee
with $\ell = - \ln (m_\ssB \e)$ diverging logarithmically when $\e
\to 0$ and $\gamma = 0.5772\dots$ being the Euler-Mascheroni
constant.

The trouble here lies in the fact that the classical solution for
the bulk field coupled to a brane diverges when evaluated at the
brane source. This is a completely generic feature for branes
having codimension 3 or larger --- {\it e.g.} the divergence of
the Coulomb field at the position of the source charge. It is also
generic for codimension 2, although exceptions in this case also
arise, such as for the conical singularities arising in the static
gravitational fields sourced by some
\cite{cosmicstrings,conicaldefects,SLED} but not all
\cite{nonconicalcod2,JJ} codimension-2 branes. And the generic
resolution to this problem lies in the need to renormalize the
brane-bulk couplings {\it even at the classical level}
\cite{GW,CdR}. As these references show (and is briefly summarized
in Appendix \ref{AppCouplRen}), the requirement that bulk $\phi$
propagators be finite implies the brane couplings also diverge
logarithmically in the limit $\e \to 0$, with the result
\be \label{ClassicalRG2}
    \lambda_2 = \frac{\bar \lambda_2}{1+
    {\bar\lambda_2\hat\ell}/{2\pi}} \quad \text{and} \quad
    \lambda_4 = \frac{\bar \lambda_4}{\( 1+ {\bar
    \lambda_2\hat\ell}/{2\pi} \)^4} \,, \nn
\ee
where the $\bar\lambda_i$ are renormalized quantities that remain
finite in the limit that $\e \to 0$, and
\be
    \hat\ell = -\ln (\mu \e) = \ell + \ln\left( \frac{m_\ssB}{\mu}
    \right) \,,
\ee
for an arbitrary renormalization scale $\mu$. For later purposes
we remark that because the term in the action involving
$\lambda_2$ is quadratic in $\phi$, it is possible to evaluate the
classical scalar propagator, including the brane-bulk mixing,
without having to assume that $\lambda_2$ or $\bar\lambda_2$ are
small (see Appendix \ref{AppCouplRen} for details). In particular
the domain of validity of eq.~\pref{ClassicalRG2} includes the
regime of large $\bar\lambda_2$.

If we regularize by replacing $\phi_0$ with $\phi_\e$, the
boundary condition which determines $\bar\phi$ becomes
\be \label{BC2config}
    - 2 \pi r \, \partial_r \phi + U_b'(\phi)
    = 2 \pi \bar\phi + U_b'(\phi_\e)
    = \left( 2 \pi - \lambda_2 \, z_\e + \lambda_4 \, z_\e^3
    \bar\phi^2 \right) \bar\phi = 0 \,.
\ee
For $\lambda_2 < 2\pi/z_\e$ this only admits the trivial solution,
$\bar\phi = 0$, but for $\lambda_2 > 2\pi/z_\e$ three solutions
are possible: $\bar\phi = 0$ and $\bar\phi = \pm \phi_c$, with
\be
    \phi_c^2 = \frac{(\lambda_2 - 2\pi/z_\e)}{\lambda_4 \, z_\e^2} \,.
\ee
Notice that the criterion distinguishing the existence of one or
three solutions depends only logarithmically on $m_\ssB$ (through
its appearance in $z_\e$), and can be phrased in a
regularization-independent manner by trading $\lambda_2$ for
$\bar\lambda_2$. In particular, the condition $\lambda_2 <
2\pi/z_\e$ ensuring only $\bar\phi = 0$ is a solution then becomes
$\bar\lambda_2 < 2\pi/c$, where $c = \ln 2 - \gamma -
\ln(m_\ssB/\mu)$ defines the finite part of $z_\e \equiv \hat\ell
+ c$.

The physical content of these expressions becomes clearer once the
relative energy of these solutions is computed using the classical
energy density, $\cH(\bar\phi)$, which is finite once it is
expressed in terms of the renormalized quantities $\bar\lambda_i$.
Explicitly, we have
\ba \label{energyDensity2}
    {\mathcal H} &=& \lim_{\e \to 0} \left\{
    2 \pi \int_\e^\infty r \d r \left[
    \frac12 (\partial_r \phi)^2 + \frac12 \, m_\ssB^2 \phi^2 \right]
    +U_b(\phi_\epsilon) \right\} \\
    &=& \lim_{\e \to 0} \left\{ \pi \bar \phi^2
    \int_{m_\ssB\e}^\infty \d z \, z
    \Bigl[ \left( K_0' \right)^2 + \left( K_0 \right)^2
    \Bigr] +U_b(\phi_\epsilon) \right\} \,.
\ea
The integral may be evaluated in closed form (see Appendix
\ref{AppBessel}), to give
\ba
    {\mathcal H} &=& \lim_{\e \to 0} \left\{
    -\frac{\pi}{2} \, \bar \phi^2 m_\ssB^2 \e^2
    K_0(m_\ssB \e) \Bigl[ K_0(m_\ssB\e)- K_2(m_\ssB \e) \Bigr]
    - \frac12 \lambda_2 \phi^2_\e + \frac{1}{4}\lambda_4
    \phi^4_\e \right\} \nn\\
    &=& \lim_{\e \to 0} \left\{
    \pi \bar \phi^2 z_\e - \frac 12 \lambda_2
    \bar \phi^2 z_\e^2 +\frac 14 \lambda_4 \bar \phi^4
    z_\e^4 + \mathcal{O}(\e) \right\}  \,,
\ea
which uses the asymptotic form $K_2(m_\ssB \e) \simeq 2/(m_\ssB
\e)^2$ for small $\e$. Using the asymptotic limit of
eq.~\pref{ClassicalRG2} for $\bar\lambda_2 \hat\ell \gg 2\pi$,
\be \label{ClassicalRG2a}
    \lambda_2 \simeq \frac{2\pi}{\hat\ell} \left[ 1
    - \left( \frac{2\pi}{\bar\lambda_2 \hat\ell} \right)
    + \cdots \right] \quad \text{and} \quad
    \lambda_4 \simeq \left( \frac{
    2\pi}{\bar\lambda_2 \hat\ell} \right)^4 \, \bar \lambda_4
    + \cdots\,, \nn
\ee
we find the finite limit
\be \label{Heff}
    \cH = \frac12 \, g_2 \, \bar\phi^2 + \frac14 \, g_4 \bar\phi^4
    \quad \text{with} \quad
    g_2 = 2\pi \left(\frac{2\pi}{\bar\lambda_2}
    - c\right) \quad\text{and}\quad
    g_4 = \left( \frac{2\pi}{\bar \lambda_2}
    \right)^4 \bar \lambda_4 \,,
\ee
where $c = \ln 2 - \gamma - \ln (m_\ssB/\mu)$, as above.

Notice the kinetic energy has combined with the bulk potential
energy to partially cancel the quadratic term in the brane
potential, with the solution $\bar\phi = 0$ being energetically
preferred for $\bar\lambda_2 < 2\pi/c$ --- the same criterion
found earlier. Notice also that $c > 0 $ if $\mu > \mu_\star =
\frac12 \, e^\gamma m_\ssB \simeq 0.89 \, m_\ssB$, and $c < 0$ if
$\mu < \mu_\star$. $c$ vanishes at the dividing case, $\mu =
\mu_\star$, at which point the quadratic term is simply
\be
    g_2 = \frac{4\pi^2}{\ls} \,,
\ee
with $\ls \equiv \bar\lambda_2(\mu_\star)$. In terms of
renormalized quantities the criterion for symmetry breaking
becomes $\ls < 0$, in which case the scalar v.e.v. is
\be \label{renvev}
    \phi_c^2 = - \frac{g_2}{g_4} = - \frac{\ls^3}{
    4\pi^2 \bar\lambda_4} \,.
\ee

These calculations illustrate how the vacuum energetics of a bulk
scalar depends crucially on the codimension of the brane to which
it is coupled. In all cases the competition between gradient and
potential energies in general allows the brane potential to drag
the bulk scalar v.e.v. away from the value which minimizes
$U_\ssB$. But in the codimension-1 case the marginal strength of
brane instability which distinguishes a nonzero from a vanishing
v.e.v., $m_b = 2m_\ssB$, depends strongly on the UV-sensitive
scale $m_\ssB$. By contrast, the corresponding criterion for
codimension-2 branes, $\bar\lambda_2 = 2\pi/c$, is comparatively
insensitive to $m_\ssB$ because it is the larger gradient energies
which replace $U_\ssB$ in dominating the fight against $U_b$. (The
situation for higher codimension is explored in Appendix
\ref{AppHigherCod}, below.)

\subsection{The Self-Localized State}
\label{sec.SelfLocalizedState}

Since we expect the quadratic term in $\cH$ to describe the mass
of small fluctuations about the background configuration, there is
a potential puzzle hidden in the weak dependence of $g_2$ on
$m_\ssB$. To see why, suppose the two extra dimensions are a
square torus of volume $V_2 = L^2$, for which in the absence of
the brane interactions we would normally expect a Kaluza Klein
spectrum to be labelled by two integers, $n_1$ and $n_2$, with
masses
\be
    M^2_{n_1 n_2} = m_\ssB^2 + M_c^2 (n_1^2 + n_2^2)
    \ge m_\ssB^2\,,
\ee
where $M_c = 2\pi/L$. The puzzle is that all of these states have
masses larger than $m_\ssB$, a result which seems hard to
reconcile with a mass governed by the size of the quadratic term,
$\frac12 \, g_2 \bar\phi^2$, of $\cH$.

We next show that the resolution of this puzzle lies in the
existence of a lower-mass `bound' state whose mass lies in the
gap, $m < m_\ssB$, and whose presence relies on the influence of
the interactions between $\phi$ and the brane. Furthermore, this
state is localized near the brane by these interactions, in the
sense that its wave-function falls exponentially away from the
brane, with a characteristic size of order $a_\ssB \sim 1/k$,
where $k^2 = m_\ssB^2 - m^2$. We call this the self-localized
state, inasmuch as its localization is a direct consequence of the
scalar-brane interactions (rather than due to a geometric effect,
like warping, such as considered in
ref.~\cite{warpedlocalscalar}).

\subsubsection*{The Fluctuation Spectrum}

To this end consider small fluctuations in the bulk scalar field,
\be
    \phi(t,r,\theta) = \varphi(r) + \Phi_{n\o}(r) e^{in\theta - i\o
    t} \,,
\ee
labelled by their energy, $\o$, and angular momentum,\footnote{We
assume here an axially-symmetric bulk, such as might be generated
(say) by two branes.} $n$. $\varphi(r)$ here denotes any of the
vacuum configurations described above. The field equation obtained
by linearizing eq.~\pref{cod2eom} in polar coordinates is
\be \label{SelfLocEq}
    \frac1r \, \partial_r \Bigl( r\, \partial_r\Phi_{n\omega} \Bigr)
    - \frac{n^2}{r^2} \, \Phi_{n\o} -k^2 \Phi_{n\o}
    = \frac{\delta_+(r)}{2 \pi r} \, \left( -\lambda_2 + 3 \lambda_4 \varphi^2
    \right) \Phi_{n\o} \,,
\ee
where $k^2 = m_\ssB^2 - \o^2$. For the purposes of identifying the
bound state we further specialize to axially symmetric modes, and
so set $n=0$.

The steps for solving for $\Phi_\o$ closely parallel those taken
above to find the background solution. Away from $r=0$ the bulk
solution is a linear combination of the modified Bessel functions,
$K_0(kr)$ and $I_0(kr)$, although in the limit $k L \gg 1$ the
admixture of $I_0(kr)$ can be made negligibly small. In this case
the background configuration is $\varphi = \bar\phi \, K_0(m_\ssB
r)$ and the fluctuation solutions are well approximated
by\footnote{Intriguingly, recasting the field equation to remove
the single-derivative term, through the redefinition $\phi =
\psi/r^{1/2}$, leads to the Schr\"odinger equation for motion of a
point particle in a $1/r^2$ potential supplemented by a
$\delta$-function at the origin. This much-studied equation is
known to exhibit the interesting phenomena of dimensional
transmutation \cite{dimtrans} and nontrivial limit cycles
\cite{braaten}.}
\be
    \Phi_\o(r) = N_\o  K_0 (k r)  \,,
\ee
with $N_\o$ an appropriate normalization constant ({\it e.g.}
$N_\o^2 = k^2/\pi$ when $k L \gg 1$). (In this notation the tower
of KK states having masses greater than $m_\ssB$ correspond to the
ordinary Bessel functions obtained when $k$ is pure imaginary.)
The eigenvalue, $k$, is obtained by imposing the boundary
condition at $r=0$, which becomes
\be \label{bdycond}
    2\pi N_\o + U_b''(\varphi) \Phi_\o(r=0)
    = \left( 2 \pi  - \lambda_2 \hat z_\e
    + 3\lambda_4 z_\e^2 \hat z_\e
    \bar\phi^2 \right) N_\o = 0  \,,
\ee
where $z_\e = - \ln(m_\ssB \e/2) - \gamma = \hat\ell + c$ is as
defined above, and $\hat z_\e$ is the same quantity with $m_\ssB
\to k$: {\it i.e.} $\hat z_\e = z_\e + \ln(m_\ssB/k)$. This
equation is to be read as being solved for $k$, leading to the
result $\hat z_\e = 2\pi/(\lambda_2 - 3\lambda_4 z_\e^2
\bar\phi^2)$, or
\ba \label{k-constraint}
    \ln \left( \frac{k}{m_\ssB} \right)
    &=& z_\e - \frac{2\pi}{\lambda_2 - 3 \lambda_4 z_\e^2 \bar\phi^2}
    = \hat\ell + c - \frac{2\pi}{\lambda_2 - 3 \lambda_4
    (\hat \ell + c)^2 \bar\phi^2} \nn\\
    &=& c -  (2\pi/\bar\lambda_2) - (3 \bar\lambda_4
    \bar\phi^2/\bar\lambda_2)(2\pi/\bar\lambda_2)^3 +
    \O(1/\hat\ell)\nn\\
    &\to& - \left( \frac{2\pi}{\ls} \right)
    \left[1 + \left( \frac{12\pi^2 \bar\lambda_4
    \bar\phi^2}{\ls^3} \right) \right] \quad
    \text{as $\e \to 0$}
    \,.
\ea
Consequently, $k = m_\ssB \, e^{-2\pi/\lambda_{2\eff}}$, or
\be \label{SelfLocalizedMass}
    \o^2 = m_\ssB^2 - k^2 = m_\ssB^2 \left[ 1 -
    e^{-4\pi/\lambda_{2\eff}} \right] \,,
\ee
where
\be \label{l2eff}
    \frac{1}{\lambda_{2\eff}} = \frac{1}{\ls} \left[ 1
    + \left( \frac{12\pi^2 \bar\lambda_4
    \bar\phi^2}{\ls^3} \right) \right] \,.
\ee
Clearly this state lives in the gap, with $\o < m_\ssB$, provided
only that $\lambda_{2\eff} > 0$, and this mass can be
hierarchically small if $\lambda_{2\eff} \gg 4\pi$ (which lies
within the domain of validity of the approximations used, as
emphasized in Appendix \ref{AppCouplRen}).

There are now two cases to consider. When $\ls > 0$ we have
$\bar\phi = 0$ and so $\lambda_{2\eff} = \ls
> 0$, showing that the self-localized state exists. In the limit
$\ls \gg 4\pi$ we find $k \simeq m_\ssB$ and $\o^2 \simeq 4\pi
m_\ssB^2 /\ls = g_2 m_\ssB^2/\pi \simeq g_2 N_\o^2$, in agreement
with the result computed from $\d^2\cH/\d\bar\phi^2$ (once care is
taken to canonically normalize the 4D scalar field).
Alternatively, when $\ls < 0$ we have $\bar\phi = \pm\phi_c$, with
$\phi_c$ given by eq.~\pref{renvev}, and so $\lambda_{2\eff} =
-\frac12 \ls > 0$. Again a bound state exists whose mass agrees
with the result, $-2 \, g_2 N_\o^2$, obtained by differentiating
$\cH(\bar\phi)$.

\section{A Self-Localized Bulk Higgs and the Hierarchy Problem}

Because the above vacuum energetics show that the expectation
value of a bulk scalar coupled to a codimension-2 (or higher
codimension) brane is less sensitive to the details of the model's
ultraviolet completion it can be used to provide a new approach to
tackling the stability issue of the hierarchy problem. This
section builds a simple illustrative example of this mechanism, in
order to get a sense of its implications.

\subsection{The Model}
\label{sec.TheModel}

The mechanism's defining assumption is that the usual Standard
Model Higgs doublet, $H(x,y)$, is located in an extra-dimensional
bulk, while all of the other Standard Model particles --- {\it
i.e.} its gauge fields, $A^a_\mu(x)$, and fermions, $\psi_k(x)$
--- reside on a brane whose codimension is at least two. (In
practice we focus on the codimension-2 case in what follows, but
generalizations to more general codimension are conceptually
straightforward.) We take the brane potential to prefer an
$SU_\ssL(2) \times U_\ssY(1)$ breaking phase, while the bulk
potential favors $SU_\ssL(2) \times U_\ssY(1)$ invariance:
\be
    U_\ssB = m_\ssB^2 \, H^* H \quad \text{and} \quad
    U_b = - \lambda_2 \, H^* H + \lambda_4 \, (H^* H)^2 \,,
\ee
where $m_\ssB^2$, $\lambda_2$ and $\lambda_4$ are all real and
positive (evaluated at $m_\ssB \e \ll 1$).

We have seen that the classical vacuum of the higher-dimensional
theory depends crucially on the sign of the renormalized coupling,
$\ls$, defined at the (large) scale $\mu_\star \simeq 0.89 \,
m_\ssB$. Notice in this regard that eq.~\pref{ClassicalRG2}
implies that both signs of $\ls$ can be consistent with positive
$\lambda_2$ when $\ell = - \ln( m_\ssB \e)$ is sufficiently large.
We take $\ls < 0$ in order to ensure that the total classical
energy is minimized by an $SU_\ssL(2) \times U_\ssY(1)$ breaking
configuration.

If we had had $SU_\ssL(2) \times U_\ssY(1)$ invariance throughout
the bulk we would at this point be able to perform a gauge
transformation to ensure that the Higgs doublet everywhere takes
the unitary gauge form, $H = \frac{1}{\sqrt2} \left( 0, \chi
\right)^\ssT$, with $\chi$ real. However because we only have
gauge invariance at the brane position this choice can only be
made at $y^m = 0$: $H_0 = \frac{1}{\sqrt2} \left( 0, \chi_0
\right)^\ssT$, where $H_0 \equiv H(x,0)$. Away from the brane $H$
in general contains 4 real fields, $H = \frac{1}{\sqrt2}
\left(\zeta_1 + i \zeta_2, \chi + i \zeta_3 \right)^\ssT$, each of
which must solve its appropriate field equations.

The arguments of the previous sections imply that the classical
vacuum solutions may be constructed in terms of $K_0(m_\ssB r)$
and $I_0(m_\ssB r)$, with the coefficient of $I_0(m_\ssB r)$
negligibly small when the extra dimensions are large compared with
$m_\ssB^{-1}$ --- {\it i.e.} $m_\ssB L \gg 1$:
\be
    \zeta_i(r) = \bar\zeta_i \, K_0(m_\ssB r) \quad \text{and}
    \quad \chi(r) = \bar\chi \, K_0(m_\ssB r) \,.
\ee
As before the normalizations, $\bar\zeta_i$ and $\bar\chi$, are
determined by the boundary conditions at $r=0$, and so
$\bar\zeta_i = 0$ due to the choice of unitary gauge at the brane,
which implies $\zeta_i(0) = 0$ there. By contrast, the arguments
of previous sections go through verbatim to imply $\bar\chi \equiv
V^2$, with
\be \label{Vvsmb}
    V^4 = - \, \frac{g_2}{g_4}
    = -\,\frac{\ls^3}{(2\pi)^2 \bar\lambda_4}
    = -\,\frac{\ls^3}{(2\pi)^3} \, M_b^4\,,
\ee
where we define $\bar\lambda_{4}/2\pi = 1/M_b^4$.

Similar arguments for the fluctuations, $\delta H$, show that in
general all four components, $\delta\zeta_i$ and $\delta\chi$, are
nonzero in the bulk. However the choice of unitary gauge at the
brane endows $\delta \zeta_i$ with the boundary condition that it
must vanish, and this in turn implies that none of these fields
localizes at the branes in the same way that $\delta \chi$ does.

Since $SU(2)\times U(1)$ is only a global symmetry in the bulk,
one might worry that its breaking by $H$ implies that the $\delta
\zeta_i$ contain KK towers of Goldstone modes that are
systematically light compared with $m_\ssB$. These could be
phenomenologically dangerous, even if their couplings must be
derivatively suppressed \cite{GB}. However (as shown in appendix
\ref{appendixGoldModes} in more detail) the only Goldstone modes
in the bulk-Higgs sector are the three self-localized states for
the fields $\delta \zeta_i$ that are eaten by the brane gauge
fields {\it via} the usual Higgs mechanism. All other states with
energies smaller than $m_\ssB$ are typically removed by the
boundary condition that requires $\delta\zeta_i$ to vanish at the
brane, leaving the lightest remaining {\it bona fide} KK modes in
$\delta\zeta_i$ with a mass of order $m_\ssB$.

\subsection{Scales and Naturalness}

We now ask how $V$ depends on the other scales in the problem, in
order to identify whether the choices required to have
sufficiently small masses for electroweak gauge bosons are
technically natural -- {\it i.e.} stable against integrating out
very heavy degrees of freedom.

The model potentially involves several scales: among which are the
compactification scale, $M_c$; the scale of extra-dimensional
gravity, $M_* \gg M_c$, (or perhaps the string scale), which
controls our neglect of gravitational physics; the scale of brane
structure,\footnote{For instance, such structure might ultimately
arise if the 3-brane were really a higher-dimensional brane
wrapped about further, smaller extra dimensions.} $\Lambda =
1/\e$, used in earlier regularizations, and so on. In principle
the UV scale, $M \gg M_c$, to which we imagine being potentially
sensitive, can be any one of these, or some other scale associated
with other types of heavy particles.

Our choices of scales are restricted by the domain of validity of
approximations used in our calculations. For instance, use of
codimension-2 branes without resolving the brane structure when
discussing the UV physics implies $\Lambda \gg M$. Ignoring (for
simplicity) the influence of the second brane on the mode
functions ({\it i.e.} dropping the admixture of $I_0(kr)$) assumes
$k \gg M_c$, where $k^2 = m_\ssB^2 - m^2$ for the self-localized
mode. Neglect (for convenience) of gravitational effects requires
both $M_* \gg M_c$, and the condition that the spacetime
curvatures generated by the configurations of interest to be small
compared with $M_*^2$. For instance if $H$ takes values of order
$V^2$ that change over distances of order $\e$, then the resulting
gradient energies do not overly gravitate if $(\partial H)^2/M_*^6
\sim (V/M_*)^4 (\Lambda/M_*)^2 \ll 1$. In what follows we assume
all of these conditions to hold. The question we ask is {\it not}
whether these hierarchies themselves are stable under
renormalization (as this would require more information, such as
specifying a stabilization mechanism for the size of the extra
dimensions), but rather whether the choices required of the Higgs
potential to obtain an acceptably small $V$ are stable against
renormalization, given the presence of these (and possibly other)
scales.

This requires an estimate of the corrections to $U_\ssB$ and $U_b$
that might arise as various kinds of heavy particles are
integrated out. Although a precise statement of this requires
specifying the theory's UV completion, some generic statements are
possible on dimensional grounds for the corrections due to
integrating out heavy particles that interact through small
dimensionless couplings. This is because if such a particle has a
large mass $M$, then its generic contribution to a coupling,
$\lambda_i$, having dimension (mass)${}^n$ is $\delta \lambda_i
\propto M^n$. According to this kind of estimate we expect
\be
    \delta m_\ssB^2 \propto M^2 \,, \quad
    \delta \lambda_2 \propto \ln M \quad\text{and}\quad
    \delta \lambda_4 \propto M^{-4} \,.
\ee
As a result it is natural to expect the corrections to $m_\ssB$ to
be dominated by the heaviest particles that can contribute, and so
generically expect $m_\ssB$ to be comparable to the largest scales
in the problem (and in particular to satisfy $m_\ssB \gg M_c$ and
$m_\ssB \gg M_\ssW$). It is the large size of these contributions
to $m_\ssB$ that underlie the usual formulation of the hierarchy
problem in 4 dimensions, because in this case the scale of the
Higgs v.e.v. turns out to be proportional to $|m_\ssB|$.

By contrast, in the 6D model of present interest we have seen that
the size of the Higgs v.e.v. is largely independent of $m_\ssB$,
depending dominantly on the dimensionless coupling $\bar\lambda_2$
and the dimensionful coupling $\bar\lambda_4$. But $\bar\lambda_2$
is dimensionless, and so tends to depend only logarithmically on
the large UV scale $M$. Potentially more dangerous is
$\bar\lambda_4/2\pi = 1/M_b^4$ since this more directly sets the
size of $V$. However this is also not UV sensitive because
corrections to it vary inversely with the relevant particle mass
on dimensional grounds, and so are dominated by the contributions
of the {\it lightest} particles, rather than the heaviest.

As stated above, we emphasize that our goal here is not to provide
an ultraviolet completion of the bulk-Higgs model, as would be
required to understand in detail the conditions necessary to
produce a large hierarchy in the first place, as this goes beyond
the scope of this paper. Our goal is instead to point out how the
introduction of Higgs bulk-brane couplings allows interestingly
different mass-dependence in low-energy observables, and to study
what this might imply for the low-energy sector.

\subsection{Higgs-induced Mass Terms}

The phenomenology of any such Higgs hinges on the form of its
couplings to observed Standard Model particles, which are assumed
in this framework to be localized on a brane.

\subsubsection*{Gauge Couplings}

At first sight it is bizarre to restrict the SM gauge fields to a
brane and yet allow a charged matter field (the Higgs doublet)
live in the bulk. This is bizarre because the $SU_\ssL(2) \times
U_\ssY(1)$ symmetry transformations are global transformations in
the bulk (since there is no spin-one field there to `gauge' them),
yet are local on the brane. Nonetheless, it must be possible
because we could imagine the UV completion of the brane of
interest being an ordinary gauge-Higgs theory containing vortex-
or domain-wall-type defects. Since the Higgs field defining the
defect typically vanishes at the interior of such a
vortex/domain-wall, there generically should be spin-1 states
which would be very massive given the nonzero Higgs in the bulk,
but which can remain light by being localized on the brane.
($D$-branes also contain localized spin-1 fields.)

More precisely, it can be shown that gauge invariance of such a
theory can always be ensured through an appropriate choice of
effective interactions (or counter-terms) on the brane. Slightly
generalizing the discussion of ref.~\cite{5Dbranegaugebulkhiggs}
to codimension two, we may see this formally by taking the Higgs
covariant derivatives to be
\be
    D_\ssM H(x,y) = \partial_\ssM H(x,y)
    - \delta^2(y) \delta_\ssM^\mu i\kappa g A_\mu^a(x) T_a H(x,y)
    \,.
\ee
Here $T_a$ are gauge generators, and as before the $x^\mu$ lie
along the brane directions while the $y^m$ are transverse. $g$
here denotes the dimensionless gauge coupling on the brane and
$\kappa$ is a dimensionful constant, required in order to counter
the dimensions of the delta function. One reason for the need for
brane counter-terms can be seen because the off-brane components
of this covariant derivative are {\it not} actually covariant
under the gauge transformation
\be
    \delta H(x,y) = \delta^2(y) i \kappa \Omega^a(x) T_a H(x,y)
    \,,
\ee
even when supplemented by the standard $x^\mu$-dependent
nonabelian transformations of $A_\mu^a(x)$. They are not because
there is no gauge potential in $D_m H = \partial_m H$ to cancel
the term arising when the derivative acts on the delta function.
There is however a counterterm that can be added on the brane such
that the entire combination {\it is} gauge invariant.

The implications of a bulk Higgs v.e.v. for gauge boson masses can
be seen by writing out the bulk and brane kinetic terms
\ba \label{bulkkin}
    \cL_{\rm kin} &=& - \int \d^2y \; D_\ssM H^* D^\ssM H
    - \kappa_b \cD_\mu H_0^* \cD^\mu H_0 \nn\\
    &=& - \int \d^2y \; \Bigl[ \partial_\ssM H^* \partial^\ssM H
    \Bigr] - (\kappa+\kappa_b) \cD_\mu H_0^* \cD^\mu H_0
    + \kappa \partial_\mu H_0^* \partial^\mu H_0 \\
    && \qquad\qquad \qquad + \frac{\kappa g^2}{2}
    [1 - \kappa \delta^2(0)] \,
    (H_0^* \left\{T_a, T_b \right\} H_0)
    \, A_\mu^a A^{b\mu} \,,\nn
\ea
where $\cD_\mu H_0 \equiv \partial_\mu H_0 -ig A_\mu^a T_a H_0$ is
the standard covariant derivative on the brane. This shows that
all of the gauge-boson mass terms appear in the brane kinetic term
provided $\kappa \, \delta^2(0) = 1$ (and so $\kappa = O(\e^2)$).
Notice that this implies $\kappa + \kappa_b \sim
\kappa_b$ for any scale $\kappa_b \gg O(\e^2)$.

Superficially the gauge-boson mass obtained from these equations
diverges as $\e \to 0$, due to the divergence there of $H_0$.
However, this divergence is countered by the renormalization of
all Higgs-brane interactions due to the generic `dressing' of
these couplings \cite{GW,CdR} by the Higgs-brane mixing,
$\lambda_2$:
\be
    \kappa_b = \frac{\bar\kappa_b}{(1 + \bar\lambda_2 \hat
    \ell/2\pi)^2} \,.
\ee
Going to unitary gauge at the brane position, for which $H_0 =
\frac{1}{\sqrt2} \, \left( 0, \chi_0 \right)^\ssT$, with $\langle
\chi \rangle = V^2 K_0(m_\ssB r)$, we then have $\kappa_b \langle
\chi_0 \rangle^2 = \bar\kappa_b V^4 (2\pi/\ls)^2$ as $\e \to 0$.

The $SU_\ssL(2) \times U_\ssY(1)$ doublet structure of the Higgs
then leads in the standard way to the prediction $M_\ssZ =
M_\ssW/\cos\theta_\ssW$, where $\theta_\ssW$ is the weak mixing
angle, and the $W$-boson mass is, $M_\ssW = \frac12 \, g v$, with
\be \label{gaugebosonmasses}
    v^2 = \left( \frac{2\pi}{\ls} \right)^2
    \bar\kappa_b V^4
    = (246 \; \text{GeV})^2 \,.
\ee
Taking $\bar\kappa_b = 1/f^2$, this shows that successful
phenomenology requires $V^2 = f v (\lsa/2\pi)$: {\it i.e.} $V$ is
the geometric mean between 246 GeV and the scale $\lsa f/2\pi$:
\be \label{phenvev}
    V \sim 10^9 \; \text{GeV} \left( \frac{\lsa f/2\pi}{10^{15}
    \; \text{GeV}} \right)^{1/2} \,.
\ee

Recall that within the present framework we have $V^4 =
|\ls/2\pi|^3(2\pi/\bar\lambda_4)$ --- {\it c.f.} eq.~\pref{Vvsmb}
--- so defining $M_b$ by $\bar\lambda_4/2\pi = 1/M_b^4$ as before
we see that eq.~\pref{gaugebosonmasses} also implies that $M_b$
must be of order
\be
    M_b \sim \sqrt{v f} \left( \frac{2\pi}{\lsa}
    \right)^{1/4}\,.
\ee \label{hprob}
This requires either $M_b \sim f \sqrt{2\pi/\lsa} \sim v$, or a
hierarchy $v \ll M_b \ll f\sqrt{2\pi/\lsa}$, if $f\sqrt{2\pi/\lsa}
\gg v$. In the absence of a symmetry which forbids a Higgs kinetic
term but allows a quartic $(H^* H)^2$ interaction on the brane,
naturalness argues we should take $f$ and $M_b$ to be the same
order of magnitude, in which case any hierarchy between $M_b$ and
$v$ must be due to $\lsa/2\pi$ being very large or very small.
Furthermore, having $f$ and $M_b$ both larger than $v$ requires
$\lsa/2\pi \lsim O(1)$.

\subsubsection*{Fermion Couplings}

Fermion masses in this picture are similarly given by Yukawa
couplings between brane-based fermions, $\psi_k$, and the bulk
Higgs doublet. In unitary gauge on the brane, $H_0 =
\frac{1}{\sqrt2} \, \left( 0, \chi_0 \right)^\ssT$, these have the
form
\be \label{basicyukawa}
    \cL_{\rm yuk} = \frac{y_{ij}}{F} \, (\overline\psi_i \psi_j)
    \chi_0 \,,
\ee
for $y_{ij}$ a collection of dimensionless Yukawa couplings, and
$F$ representing an appropriate ultraviolet scale. The resulting
fermion masses are
\be \label{fermionmasses}
    m_{ij} = \frac{y_{ij} }{F} \, \langle \chi_0 \rangle
    = \frac{2\pi \bar y_{ij} \,V^2}{\ls F}
\ee
with
\be
    y_{ij} = \frac{\bar y_{ij}}{1 + \bar\lambda_2 \hat\ell/(2\pi)}
    \,,
\ee
being the renormalized Yukawa coupling, as required to counter the
divergence of $H$ at the brane position, and the second equality
in eq.~\pref{fermionmasses} uses $y_{ij} \langle \chi_0 \rangle =
2\pi \bar y_{ij} V^2/\ls$ in the limit $\e \to 0$, where $\langle
\chi \rangle = V^2 K_0(m_\ssB r)$.

Since $\cL_{\rm yuk}$ breaks flavor symmetries --- unlike the
Higgs kinetic terms --- the scale $F$ need not be of the same
order of magnitude as\footnote{This could arise, say, if the
3-brane is really a higher-dimensional brane wrapped in extra
dimensions, and the flavor structure is associated with this
wrapping, since this would suggest $F \simeq \Lambda \gg f$.} $f$.
In particular, since the dominant contributions to couplings
having dimensions of inverse mass come from the lightest scales to
contribute, $F$ is typically set by the smallest UV scale which
involves flavor-violating physics while $f$ can be much smaller
than this. Because of this eqs.~\pref{gaugebosonmasses} and
\pref{fermionmasses} may contain the seeds of an explanation of
the observed smallness of most fermion masses relative to those of
the electroweak gauge bosons, since
\be
    \frac{m_{ij}}{M_\ssW} = \frac{\bar y_{ij}}{g} \left(
    \frac{2f}{F} \right) \,.
\ee
Even a mild hierarchy, $F \gg f$, removes some of the burden of
having to require $\bar y_{ij}/g$ to be very small.

\subsection{Couplings to the Higgs Fluctuations}

We have seen that the spectrum of fluctuations in the Higgs field
generically contains an assortment of KK modes, many of whose
masses start above a large gap, $m_{\ssK\ssK} \gsim m_\ssB$. For
$m_\ssB$ sufficiently large these modes need not play an important
role in low-energy observables. The two exceptions to the above
statement are the bulk Goldstone modes, whose masses are
generically of order $M_c$, and the self-localized state whose
mass can lie within the gap below $m_\ssB$, and be hierarchically
smaller if $\lsa \gg 2\pi$. Furthermore, this latter state is
present regardless of whether or not the Higgs v.e.v. is nonzero.
These light states are likely to be the ones relevant to Higgs
phenomenology in Bulk Higgs models, and so this section computes
their couplings.

\subsubsection*{The bulk Goldstone modes}

The simplest couplings to compute are those of the bulk Goldstone
modes, $\delta\zeta_i$, because their vanishing at the brane
position guarantees they completely drop out of any brane
couplings that depend only on $H_0$ or $\partial_\mu H_0$, and not
on off-brane derivatives like $\partial_m H_0$. In particular this
ensures their removal (in unitary gauge) from the fermion Yukawa
couplings and gauge couplings described above.

\subsubsection*{The self-localized state}

Normalizing the wave-function of the self-localized state in the
extra dimensions gives a canonically normalized 4D state $h$,
where $\chi = h(x) N_\o K_0(kr)$, so $y_{ij} \chi =
(2\pi/\ls)(k/\sqrt\pi) \bar y_{ij} h$, with $k^2 = m_\ssB^2 -
m_h^2$. The couplings of $h$ to fermions are then given by
interactions of the form
\be \label{4Dlocalizedcouplings}
    \cL_{4D} = \frac{2\pi\bar y_{ij}}{\ls}
    \left( \frac{k}{{\sqrt\pi} F} \right) \,
    (\overline \psi_i \psi_j) h
    \,,
\ee
leading to dimensionless `physical' Yukawa couplings of order
\be
    \hat y_{ij} = \frac{2\pi\bar y_{ij}}{\ls}
    \left( \frac{k}{{\sqrt\pi} F} \right)
    = y_{ij}^{\rm sm} \left( \frac{m_\ssB}{\sqrt\pi f} \right)
    \left( \frac{2\pi}{\ls}
    \right)
     e^{- 4\pi /|\ls|} \,, \ee
where the argument of the exponential assumes $\ls < 0$ (as
required for a nonzero Higgs v.e.v.), and the last equality
compares to what would be expected in the SM:
\be
    y^{\rm sm}_{ij} \equiv \frac{m_{ij}}{v}
    = \bar y_{ij}\,
    \left(\frac{f}{F} \right) \,.
\ee
Notice that the quantity $(2\pi / |\ls| ) \exp [ -4\pi / |\ls|]$
falls to zero for large and small $|\ls|$, taking the maximum
value of 0.18 when $|\ls|/2\pi = 2$.

These expressions show that the self-localized Higgs couplings,
$\hat y_{ij}$, can differ significantly from what would be
expected in the SM, for two reasons. First, $\hat y_{ij}$ can be
larger than $y^{\rm sm}_{ij}$ if $m_\ssB \gg f$, and if
sufficiently large the self-localized state becomes a strongly
coupled broad resonance. Second, $\hat y_{ij}$ can differ from
$y^{\rm sm}_{ij}$ because of its dependence on $\ls/2\pi$, which
acts to suppress $\hat y_{ij}/y^{\rm sm}_{ij}$ in the limit that
$\lsa/2\pi$ is either very large or very small. This possibility
of having $\hat y_{ij}$ differ from the SM expectation contrasts
with 4D intuition based on the couplings of a single scalar whose
v.e.v. generates mass, since such a scalar must have couplings
given by the ratio $m_{ij}/v$. The reason this conclusion does not
hold in the extra-dimensional case is that because the v.e.v.,
$\langle H(x,0) \rangle$, responsible for generating masses
generically receives contributions from many KK modes and not just
the v.e.v. of the single 4D self-localized state, $h$.

\section{Possible Signatures of a Bulk Higgs Scenario}

We next sketch some of the qualitative signatures and constraints
that might be expected for the kind of Higgs scenario described
above. What is to be expected depends somewhat on the choices made
for the various scales in the problem, so we divide the discussion
according to four simple options according to whether or not we
take $\lsa$ to be large or small, and whether we take $M_c \sim 1$
TeV, or $M_c \sim 10^{-2}$ eV (as for large-extra-dimensional
models).

\subsection{Inclusive Processes}

We first consider inclusive processes for which a specific Higgs
state is not measured, and so which involve a summation over all
possible KK modes. These are largely insensitive to the specifics
of individual modes, such as the details of the self-localized
state.

\subsubsection*{Fermion-fermion scattering}

An important inclusive observable is the rate for fermion-fermion
scattering mediated by a virtual Higgs.  The amplitude for this
process is of order
\be
    \cA(\psi_i \psi_j \to H \to \psi_r \psi_s)
    \simeq \frac{y_{ij} \, y_{rs}}{F^2} \, i G_{p}(0;0)
    \; \delta^4(p_i + p_j - p_r - p_s) \,,
\ee
where $p^\mu \equiv (p_i + p_j)^\mu = (p_r + p_s)^\mu$. Here
$G_p(y;y')$ is the bulk Higgs propagator, Fourier transformed in
the brane directions, $x^\mu$, but evaluated in position space in
the off-brane directions, $y^m$. $G_p(0;0)$ denotes the same
quantity evaluated at the brane position, and is given (see
Appendix \ref{AppCouplRen} for details) in terms of the
corresponding propagator in the absence of brane-Higgs couplings,
$D_p(y;y')$, by
\be
    G_p(0;0) = \frac{D_p(0;0)}{1 - i\lambda_2 D_p(0;0)} \,.
\ee
Eliminating $y_{ij}$, $y_{rs}$ and $\lambda_2$ in terms of the
renormalized quantities, $\bar y_{ij}$, $\bar y_{rs}$ and
$\bar\lambda_2$, and taking $\e \to 0$, we find the finite result
\ba
    \cA(\psi_i \psi_j \to H \to \psi_r \psi_s)
    &\simeq& \frac{\bar y_{ij} \, \bar y_{rs}}{\bar\lambda_2 F^2}
    \left[ \frac{1}{1- i\bar\lambda_2 D^\mu_p(0;0)} \right]
    \; \delta^4(p_i + p_j - p_r - p_s) \nn\\
    &\simeq& \frac{y^{\rm sm}_{ij} \, y^{\rm sm}_{rs}}{\bar\lambda_2 f^2}
    \left[ \frac{1}{1- i\bar\lambda_2 D^\mu_p(0;0)} \right]
    \; \delta^4(p_i + p_j - p_r - p_s) \,,
\ea
where $iD_p^\mu(0;0) = (1/2\pi) \ln(\mu/P)$, where $P^2 = p^2 +
m_\ssB^2$.

If this same process were computed using the exchange of a massive
4D SM Higgs scalar, we'd have instead obtained
\be
    \cA^{\rm sm}(\psi_i \psi_j \to H \to \psi_r \psi_s)
    \simeq y^{\rm sm}_{ij} \, y^{\rm sm}_{rs}
    \left[ \frac{1}{p^2 + m_\ssH^2} \right]
    \; \delta^4(p_i + p_j - p_r - p_s) \,,
\ee
and so the leading effect is to replace the scale $p^2 + m_\ssH^2$
by $\bar\lambda_2 f^2 [1-i\bar\lambda_2 D_p^\mu(0;0)]$. The
absence of an observed signal therefore implies the
order-of-magnitude bound
\be
    \bar\lambda_2 f^2 \left[1 + \frac{\bar\lambda_2}{2\pi} \ln
    \left( \frac{P}{\mu} \right) \right] \gsim O(100 \;
    \text{GeV} )^2 \,,
\ee
where $P^2 = (p_i + p_j)^2 + m_\ssB^2 = (p_r + p_s)^2 + m_\ssB^2$.

If reactions of this type were to mediate flavor-changing neutral
currents, the strong restrictions on these could potentially bound
the scale $F$ to be quite large. However, because the Yukawa
couplings can have the same flavor structure as in the SM, there
can be a GIM mechanism at work \cite{GIM} that naturally
suppresses the dangerous flavor-changing neutral current (FCNC)
reactions produced by bulk-Higgs exchange. We henceforth assume
this to be true, and therefore do not further worry about bounds
on the fermion couplings due to FCNCs.

\subsubsection*{Vacuum Polarization}

As is well known, the contributions to loops of the SM Higgs is
well constrained by precision electroweak measurements. The main
source of these contributions is through the Higgs contribution to
the vacuum polarization of the electroweak gauge bosons. For an
extra-dimensional bulk Higgs, this contribution is of order
\ba \label{vacpol}
    \Pi_{ab}^{\mu\nu}(p) &\simeq& g^2 \kappa_b^2 \, \text{tr}(T_aT_b)
    \int \frac{\d^4q}{(2\pi)^4}
    (2p-q)^\mu (2p-q)^\nu \, iG_q(0;0) \, iG_{p-q}(0;0) \nn\\
    &\simeq& \frac{g^2 \bar\kappa_b^2}{\bar\lambda_2^2}
    \, \text{tr}(T_aT_b)
    \int \frac{\d^4q}{(2\pi)^4} \left[ \frac{(2p-q)^\mu
    (2p-q)^\nu}{[1 - i \bar\lambda_2 D^\mu_p(0;0)]
    [1 - i \bar\lambda_2 D^\mu_{p-q}(0;0)]} \right] \,,
\ea
plus a possible tadpole term. Since the remaining integration,
$\d^4q$, diverges in the ultraviolet it must be regularized, and
this is most conveniently done using dimensional regularization.

Of most interest for phenomenological purposes is the contribution
to the oblique parameters $S$, $T$ and $U$ \cite{SM,oblique},
which involve those terms in $\Pi^{\mu\nu}_{ab}$ having the tensor
structure $(p^2 \eta^{\mu\nu} - p^\mu p^\nu)$. Since the Higgs is
an $SU_\ssL(2) \times U_\ssY(1)$ doublet, it automatically
preserves the accidental custodial $SU_c(2)$ symmetry
\cite{SM,custodial} that preserves the successful mass relation
$M_\ssW = M_\ssZ \cos\theta_\ssW$, thereby suppressing its
contribution to $T$ and making $S$ of most interest. Because all
mass dependence in eq.~\pref{vacpol} is logarithmic, recalling the
definition $\bar\kappa_b = 1/f^2$ and extracting the conventional
factors of $g^2/4\pi$, we obtain the estimate
\be
    S \sim \frac{1}{4\pi \bar\lambda_2^2}
    \left( \frac{p^4}{f^4}
    \right) \,,
\ee
where $p^2$ represents the momentum transfer of interest. Applied
to LEP experiments we may take $p^4 = M_\ssZ^4$ and $|S| < 0.1$ to
conclude $\bar\lambda_2 f^2 \gsim v^2$.

\subsection{Higgs Decays to Fermions}

Another class of observables involve specifying a specific Higgs
KK mode. Perhaps the simplest of these is the decay rate for
specific Higgs states into SM particles (although this decay need
not dominate the lifetime of a given KK mode because it must also
compete with other channels, such as off-brane decays into the
Goldstone modes $\delta\zeta_i$).

\subsubsection*{Generic KK states}

For simplicity we start with the decay of a generic KK mode into
brane fermions, assuming the KK wave-functions, $\Psi(y)$, extend
throughout much of the extra-dimensional bulk so that $|\Psi(0)|^2
\simeq 1/V_2 \simeq M_c^2$. Once excited, such a heavy state can
decay through the interaction \pref{basicyukawa}, with the rate
\be \label{decayrate}
    \Gamma(\chi \to \bar\psi_i\psi_j) \simeq |\Psi(0)|^2
    \, \frac{|y_{ij}|^2}{F^2} \, M_\chi
    \simeq |y_{ij}|^2 \left( \frac{M_c}{F} \right)^2 M_\chi\,,
\ee
where $M_\chi \ge m_\ssB$ is the mass of the decaying mode.
(Recall that the bulk Goldstone modes, $\delta\zeta_i$, do not
decay in this way because of the requirement that they vanish at
the brane.) We see that $\Gamma \ll M_\chi$ naturally follows from
the smallness of the quantities $y_{ij}$ and $M_c/F$ (the latter
of which is particularly small in the case of large extra
dimensions). Whether these are the dominant decay channels depends
on the availability of light states in the bulk (or on other
branes) into which competing decays can proceed, and how
efficiently these Higgs decays occur.

\subsubsection*{The self-localized state}

Notice that $y_{ij}$ vanishes, strictly speaking, when $\e \to 0$
with $\bar y_{ij}$ and $\bar\lambda_{2}$ held fixed (making
eq.~\pref{decayrate} vanish logarithmically in this limit). The
same is not true of the self-localized state, whose wave-function
also diverges logarithmically at the position of the brane as $\e
\to 0$. In this case the decay rate can be computed using the
interaction of eq.~\pref{4Dlocalizedcouplings}, leading (on
neglect of final-state fermion masses) to the standard 4D
expression
\be
    \Gamma(h \to \bar \psi_i \psi_j)
    = \frac{1}{8\pi} \left| \hat y_{ij}
    \right|^2 \, m_h = \Gamma^{\rm sm}(h \to \bar \psi_i \psi_j)
    \left(\frac{m_\ssB^2}{\pi f^2} \right)
    \left( \frac{2\pi}{\ls} \right)^2
    e^{-16\pi/|\ls|} \,,
\ee
which remains nonzero as $\e \to 0$. This drops dramatically, as
required in the unlocalized limit, as $|\ls| \to 0$, and scales as
$\Gamma^{\rm sm} \, m_h^4/(16\pi m_\ssB^2 f^2)$ when $|\ls| \gg
2\pi$.

\subsection{TeV-Scale Compactifications}

Suppose, first, the compactification scale, $M_c$, lies in the TeV
range and so, in the absence of significant warping, the 4D Planck
scale, $M_p \sim M_*^2/ M_c$, comes out right if $M_* \sim
10^{10}$ GeV. This leaves lots of room to choose the other scales
of interest to be much smaller than $M_*$ in order to justify our
neglect of gravitational interactions. We do not speculate as to
how the extra-dimensional size is stabilized at this scale.

Choosing $M_c$ this large also ensures that this is the mass of
the lightest KK mode of the bulk Goldstone bosons, $\delta
\zeta_i$, ensuring that these modes do not play much of a
phenomenological role until energies are reached -- at the LHC,
God willing -- that allow the direct production of KK excitations.
The same is true of the generic KK modes of the field $\chi$,
provided we also choose $m_\ssB$ to be large enough.

We have seen that the absence of Higgs detection in oblique or in
2-fermion to 2-fermion processes implies us to choose $f
\sqrt{\lsa}$ to be at least several hundred GeV, whereas our use
of a 6D calculational framework requires both $f$ and $M_b \sim
\sqrt{vf} (2\pi/\lsa)^{1/4}$ to be $\gsim M_c$. There are then two
subcategories to consider, depending on the size of $\lsa/2\pi$.

\subsubsection*{Weak localization}

Consider first the limit of small $|\ls|$, for which $m_h \to
m_\ssB$ and $k \to 0$. Because $k$ is small, the `bound' state is
not strongly localized relative to generic extra dimensional
scales, and the breakdown of the approximation $k \gg M_c$ demands
we go beyond the simple large-volume limits used above for the
scalar v.e.v. and wave-function. Taking $\lsa \sim 0.01$ for
illustrative purposes, we see that requiring $f > M_c \sim$ few
TeV automatically ensures $f \sqrt{\lsa} \gsim $ several hundred
GeV, and so is large enough to avoid the phenomenological bounds.

For weak localization, the exponential suppression of $\hat
y_{ij}$ for small $\lsa$ allows us to choose $m_\ssB$ to be much
larger than $f$ without the Higgs-fermion couplings becoming
strong. However we cannot have all $\chi$ states be too much
higher than the TeV scale without there being a breakdown of the
low-energy effective theory, such as through the development of
unitarity problems in the scattering of longitudinal $W$ bosons
that the SM would suffer in the absence of a low-energy Higgs
particle \cite{SM,unitaritybound,usesandabuses}, and this puts an
upper bound on how large $m_\ssB$ can be. In this case the $\chi$
spectrum resembles the usual intuition for bulk fields in the
absence of brane couplings, consisting of a tower of Higgs KK
modes starting above the gap at $m_\ssB$. Furthermore, because
these particles are likely to have a significant decay rate into
the lighter bulk Goldstone states, any observed Higgs is likely to
have a significant invisible width.

Because $m_\ssB$ cannot be made exceedingly large without running
into troubles, and because $M_c$ is typically smaller, it should
be possible to observe some of the Higgs KK states at the LHC.
Although the mass-$M_c$ Goldstone states cost less energy, they
are more difficult to produce because of the absence of direct
couplings to the initial brane-based SM particles. The most likely
channel for doing so is the virtual excitation of KK modes of the
bulk state $\chi$. Convincing evidence for these Goldstone states
together with an absence for KK modes for the electroweak gauge
bosons would provide the smoking gun for this scenario: with the
Higgs in the bulk but gauge interactions localized to live only on
the branes.

\subsubsection*{Strong localization}

In the opposite limit, $|\ls| \gg 2\pi$, the lowest energy state
becomes localized to the brane with $k \simeq m_\ssB$, and its
mass drops to $m_h^2 \simeq 8\pi m_\ssB^2 / |\ls| \ll m_\ssB^2$.
In this case $m_\ssB$ can be higher than it could for weak
localization, provided that the self-localized state is lighter
than a few TeV and so can unitarize the scattering of longitudinal
gauge boson modes.

An upper limit to how large $m_\ssB$ can be is found from the
condition that this light, localized Higgs state be weakly coupled
\be
    \left| \frac{\hat y_{ij}}{y^{\rm sm}_{ij}} \right|^2
    \simeq \frac{8}{\pi} \left( \frac{m_\ssB}{f}\right)^2
    \left| \frac{2\pi}{\ls}\right|^3
    \simeq \frac{1}{8\pi} \left( \frac{m_h}{f}\right)^2
    \left( \frac{m_h}{m_\ssB}\right)^4 \,.
\ee

Large $\lsa$ also implies that the condition $f > M_c$
automatically ensures the validity of the phenomenological limits
that require $f\sqrt{\lsa}$ to be larger than several hundred GeV,
and makes the strongest constraint on $f$ the theoretical
condition that $M_b$ be larger than $M_c$.

For instance for moderately large $\lsa/2\pi \sim 10^2$, then
keeping $m_h$ at the TeV scale requires $m_\ssB \simeq 10$ TeV,
and taking $M_b \sim 1$ TeV then implies $f \sim 10$ TeV. By
contrast, if $m_\ssB$ should be the largest scale considered so
far, $m_\ssB \sim M_* \sim 10^{10}$ GeV, then $\lsa/2\pi \sim
10^{20}$, and so $M_b \sim 10^{-5} f > 1$ TeV implies a strong
hierarchy between $M_b$ and $f > 10^5$ GeV whose naturality would
have to be explained. Notice that the physical couplings, $\hat
y_{ij}$, are much smaller than for the SM given these scales.

In this case $m_\ssB$ could easily be large enough to preclude the
direct detection of a Higgs KK spectrum, even at the LHC, leaving
the burden of Higgs physics carried by the single self-localized
Higgs state. In principle this can be distinguished from a SM
Higgs in several ways. First, it could well have a large invisible
width, if the mass of the self-localized state is sufficiently
large compared with the mass, $M_c$, of the bulk Goldstone modes.
Second, it can be distinguished by identifying the difference in
the strength of its couplings to fermions from those expected in
the SM.

\subsection{Large Extra Dimensions}

An alternative choice \cite{ADD,SLED,MSLED} would put the scale of
extra-dimensional gravity at $M_* \sim 10$ TeV, which then
requires $M_c \sim 10^{-2}$ eV. As a result, the upper bound
$m_\ssB < M_*$ automatically keeps the generic Higgs KK modes
light enough to potentially be seen at the LHC, yet absence of the
detection of Higgs KK modes also implies $m_\ssB$ cannot be much
below the TeV scale.

An automatic consequence of having $M_c$ so small is to make the
bulk Goldstone states, $\delta\zeta_i$, essentially massless. This
ensures that they are always kinematically available as final
states for $\chi$ decays, making a significant invisible width for
this state inevitable. In fact, the very lightest KK Goldstone
modes in this scenario are light enough to mediate forces between
macroscopic bodies, with generically near-gravitational strength,
making them potentially relevant to precision tests of Newton's
inverse-square law for gravity. Their presence is nonetheless
unlikely to have been already ruled out due to the absence of
direct couplings to brane matter, and the derivative nature of
their Goldstone interactions.

In this scenario the conditions $f, M_b \gsim M_c$ pose no
significant constraint, with more information coming from the
phenomenological conditions that $f\sqrt{\lsa}$ be larger than a
few hundred GeV. Notice that if we also require $f \lsim M_*$ then
we must have an upper bound $\lsa \lsim 10^4$, and so the
self-localized state cannot be more than a few orders of magnitude
lighter than $m_\ssB$.

Because the KK tower of modes is so narrowly spaced -- by $O(M_c)$
-- they provide almost a continuum of states. Although each of
these modes couples with gravitational strength, their phase space
makes their inclusive production cross section of order the
weak-interaction size \cite{ADD}. Once the Higgs is produced, its
phenomenology is likely to resemble that of extra-dimensional
gravitons \cite{ADDphen} or other bulk matter fields
\cite{SLEDphen}, including likely large invisible decay channels.

\section{Conclusions}

In this paper we examine a new way for brane-world scenarios to
change how we think about low-energy naturalness problems. We do
so by showing how oft-neglected couplings to branes can
dramatically change the vacuum energetics and low-energy spectrum
for bulk scalar fields. In particular, we show that when coupled
to codimension-2 branes bulk scalar fields can have two unusual
properties:
\begin{itemize}
\item They can acquire v.e.v.s that are only logarithmically
related to the size of the UV-sensitive quadratic term, $\frac12
m_\ssB^2 \phi^2$, in the bulk Higgs potential;
\item They can acquire low-energy KK modes that are localized to
the branes (without the need for warping), and whose mass can lie
inside the naive gap below the energy set by the mass scale
$m_\ssB$.
\end{itemize}

We further use these two observations to explore the possibility
of building phenomenological brane-world models for which all
Standard Model particles (save the Higgs) are trapped on a brane,
but with the Higgs allowed to live in the bulk. We estimate the
size of the effective couplings of such a Higgs to gauge bosons
and fermions on the brane, and use these to estimate the sizes of
masses and couplings to the Higgs KK modes.

We do not try to identify ultraviolet completions of the
bulk-Higgs model, and so do not identify at a microscopic level
why the electroweak hierarchy exists in the first place. Our focus
is instead on whether such a hierarchy can be technically natural
purely within the low-energy theory. We identify in
eq.~\pref{hprob} the main obstacle to systematically raising the
UV scale of this effective theory above the weak scale, since this
equation generically requires the two dimensionful parameters $f$
and $M_b$ --- governing the size of the brane potential term
$(H^*H)^2/M_b^4$ and the brane kinetic term $(D_\ssM H^* D^\ssM
H)/f^2$ --- either to satisfy $M_b \sim f
\sqrt{2\pi/\lambda_{2\star}}$ with both near the electroweak
scale, or to satisfy the hierarchy $M_b \ll f
\sqrt{2\pi/\lambda_{2\star}}$ if both are large compared with the
electroweak scale. This latter hierarchy shows how the problem
gets recast with a bulk Higgs, since both interactions are allowed
by the same symmetries, making it unnatural for them to have
coefficients suppressed by very different scales.

We provide a very preliminary discussion of possible signals and
constraints on these models, including the observation that most
realizations predict a significant invisible width for any
observed `Higgs', once detected. Simple estimates are made of
Higgs decay rates into SM particles, the scattering rate for
fermions due to virtual Higgs exchange, and the contribution of
virtual Higgs loops to gauge boson vacuum polarization. These are
used to outline the qualitative features of Higgs phenomenology
within this class of models. In all cases we find that the
phenomenology of these models is sufficiently interesting to bear
further, more detailed study.

\section*{Acknowledgements}

We wish to thank Georges Azuelos, Massimo Giovannini, Quim Matias
and James Wells for helpful comments and suggestions. This
research has been supported in part by funds from the Natural
Sciences and Engineering Research Council (NSERC) of Canada. CB
also acknowledges support from CERN, the Killam Foundation, and
McMaster University. CdR is funded by an Ontario Ministry of
Research and Information (MRI) postdoctoral fellowship. Research
at the Perimeter Institute is supported in part by the Government
of Canada through NSERC and by the Province of Ontario through
MRI.

\appendix

\section{Some Properties of Bessel Functions}
\label{AppBessel}

This appendix summarizes a few properties of modified Bessel
functions which are used in the main text. The modified Bessel
functions are linearly independent solutions to the differential
equation
\be
    z^2 y'' + z y' - (z^2 + \nu^2) y = 0 \,,
\ee
with $I_\nu(z)$ chosen to be regular at $z=0$ and $K_\nu(z)$
chosen to fall off to zero as $z \to \infty$. They are defined in
terms of ordinary Bessel functions, $J_\nu(z)$, and Hankel
functions, $H_\nu^{(1)}(z)$, by
\be
    I_\nu(z) = i^{-\nu} J_\nu(iz) \quad\text{and}\quad
    K_\nu(z) = \frac{\pi}{2} \, i^{\nu+1} H_\nu^{(1)}(iz) \,.
\ee
The expansion of these functions for small argument is used in the
text. For $0 < z \ll \sqrt{\nu+1}$ it is given by
\be
    I_\nu(z) \simeq \frac{1}{\Gamma(\nu+1)} \left( \frac{z}{2}
    \right)^\nu \,, \quad
    K_0(z) \simeq - \ln \left( \frac{z}{2} \right) - \gamma
    \quad \hbox{and} \quad
    K_\nu(z) \simeq \frac{\Gamma(\nu)}{2} \left( \frac{2}{z}
    \right)^\nu \quad \hbox{if} \quad \nu > 0 \,.
\ee
The asymptotic form at large $z$ is similarly given (for $z \gg
\left| \nu^2 - \frac14 \right|$) by
\be
    I_\nu(z) \simeq \frac{1}{\sqrt{2\pi z}} \; e^z
    \quad \text{and} \quad
    K_\nu(z) \simeq \sqrt{\frac{\pi}{2z}} \; e^{-z} \,.
\ee

The energy integral encountered in the main text can be evaluated
explicitly, using the following Bessel-function identities
\be
    K_\nu' = - K_{\nu - 1} - \frac{\nu K_\nu}{z}
    = - K_{\nu + 1} + \frac{\nu K_\nu}{z} \,,
\ee
which imply in particular $K_0' = - K_1$, $K_1' = - K_0 - K_1/z =
- K_2 + K_1/z$ and $K_2' = - K_1 - 2K_2/z$. Repeated application
of these shows that
\be
    \frac{\d}{\d z} \left[ \frac12 \, z^2 \Bigl( K_0^2 - K_1^2
    \Bigr) \right] = z \, K_0^2 \quad \hbox{and} \quad
    \frac{\d}{\d z} \left[ \frac12 \, z^2 \Bigl( K_1^2 - K_0 \, K_2
    \Bigr) \right] = z \, K_1^2 \,,
\ee
and so
\be
    z \, \Bigl( K_0^2 + K_1^2 \Bigr)
    = \frac{\d}{\d z} \left[ \frac12 \, z^2 K_0 \Bigl( K_0 - K_2
    \Bigr) \right] \,.
\ee

\section{Classical Divergences in Brane Couplings}
\label{AppCouplRen}

This appendix summarizes the derivation of the renormalization of
the codimension-2 couplings encountered in the text, with an
emphasis on identifying its domain of validity.

Consider to this end the following bulk-brane quadratic action for
a single real scalar field,
\be
    S = - \frac12 \int \d^4x \d^2y \; \Bigl[ \partial_\ssM \phi
    \, \partial^\ssM \phi + m_\ssB^2 \phi^2 \Bigr]
    + \frac12 \int d^4x \; \lambda_2 \phi^2 \,.
\ee
(The unusual sign for the brane term is chosen to be consistent
with its use in the main text.) The exact propagator,
$G(x,y;x',y')$, for this theory satisfies the differential
equation
\be
    \Bigl[ \partial_\ssM \partial^\ssM - m_\ssB^2
    + \lambda_2 \delta^2(y) \Bigr] G(x,y;x',y')
    = i\delta^4 (x-x') \delta^2(y-y') \,,
\ee
while the propagator in the absence of the brane coupling,
$D(x,y;x',y')$, instead satisfies
\be
    \Bigl[ \partial_\ssM \partial^\ssM
    - m_\ssB^2 \Bigr] D(x,y;x',y')
    = i\delta^4 (x-x') \delta^2(y-y') \,.
\ee

It is useful to regard these as the position-basis representation
of two abstract operators, $G$ and $D$, so that $G(x,y;x',y') =
\langle x,y| G | x',y' \rangle$ (and similarly for $D$). In this
case the above relations can be written $G^{-1} = D^{-1} - iV$,
where $\langle x,y | V | x', y' \rangle = \lambda_2 \delta^2(y)
\delta^4(x-x') \delta^2(y-y')$. Multiplying on the left by $D$ and
on the right by $G$ then allows this to be written as $G = D + i
DVG$, whose position-basis expression is equivalent to the
integral equation
\be \label{integraleqpos}
    G(x,y;x',y') = D(x,y;x',y') + i \lambda_2
    \int \d^4\hat x \;
    D(x,y;\hat x,0) G(\hat x,0;x',y') \,.
\ee
After Fourier transforming the translation-invariant $x^\mu$
directions
\be
    G(x,y;x',y') = \int \frac{\d^4p}{(2\pi)^4} \;
    G_p(y;y') \, e^{ip\cdot (x-x')} \,,
\ee
eq.~\pref{integraleqpos} becomes the exact statement
\be \label{integraleqmom}
    G_p(y;y') = D_p(y;y') + i \lambda_2 \,
    D_p(y;0) G_p(0;y') \,.
\ee
Since this no longer involves convolutions it may be solved
explicitly. Specializing first to $y = 0$ implies $G_p(0;y') =
D_p(0;y')/[1 - i\lambda_2 D_p(0;0)]$, which when re-substituted
into eq.~\pref{integraleqmom} gives
\be \label{integraleqmomsolved}
    G_p(y;y') = D_p(y;y') + i \lambda_2 \,
    \frac{D_p(y;0) D_p(0;y')}{1 - i\lambda_2 D_p(0;0)} \,.
\ee
Notice that no approximations have been made that implicitly
restrict us to small $\lambda_2$.

The problem with the solution, eq.~\pref{integraleqmomsolved}, is
that the quantity $D_p(0;0)$ diverges, and this observation lies
at the root of the need for renormalization. The expression for
$D_p(y;y')$ may be explicitly constructed as the following mode
sum, using polar coordinates $\{ y^m \} = \{r, \theta \}$ in the
transverse dimensions, with $r=0$ representing the brane position:
\be
    D_p(r,\theta;r',\theta') = -i \sum_{n=-\infty}^\infty
    e^{in(\theta - \theta')} \int_0^\infty \left(
    \frac{q\d q}{2\pi} \right)
    \; \frac{1}{p^2 + q^2 + m_\ssB^2} \, J_n(qr) \, J_n(qr') \,,
\ee
where\footnote{The generalization of this expression to the case
where the transverse geometry has a conical defect at the brane
position is given in ref.~\cite{CdR}.} $p^2 = p_\mu p^\mu$. To
isolate the divergence in $D_p(0;0)$ evaluate at $r = r' = 0$ and
use $J_n(0) = \delta_{n0}$ to get
\be
    D_p^\Lambda(0;0) = -i \int_0^\Lambda \left(
    \frac{q\d q}{2\pi} \right)
    \; \frac{1}{p^2 + q^2 + m_\ssB^2}
    = -\frac{i}{2\pi} \, \ln\left( \frac{\Lambda}{P} \right)
    + O\left( \frac{P^2}{\Lambda^2} \right) \,,
\ee
where $P^2 = p^2 + m_\ssB^2$.

Renormalization may also be performed without resorting to an
expansion in powers of $\lambda_2$. The goal is to redefine
$\lambda_2 = \bar\lambda_2(\Lambda) \to \bar\lambda_2(\mu)$ in
such a way as to absorb the divergence in $D^\Lambda(0;0)$:
\be
    \frac{\lambda_2}{1 - i\lambda_2 D^\Lambda_p(0;0)}
    \equiv \frac{\bar\lambda_2(\mu)}{1 - i\bar\lambda_2(\mu)
    D^\mu_p(0;0)} \,,
\ee
or, equivalently
\be
    \frac{1}{\bar\lambda_2(\Lambda)}
    \equiv \frac{1}{\bar\lambda_2(\mu)} +i \Bigl[
    D^\Lambda_p(0;0) - D^\mu_p(0;0) \Bigr]
    = \frac{1}{\bar\lambda_2(\mu)} + \frac{1}{2\pi}
    \ln\left( \frac{\Lambda}{\mu} \right) \,,
\ee
in agreement with the usage in the main text.

\section{Higher codimension}
\label{AppHigherCod}

In this appendix we examine how the arguments of $\S2$ change for
a Higgs living in a ($4+n$)-dimensional bulk coupled to a
codimension-$n$ brane, with $n \ge 3$.

We divide the discussion into a derivation of how the brane
couplings renormalize in arbitrary codimension, and then examine
the energy density that governs the size of the resulting scalar
expectation value.

\subsection{Coupling renormalization}

We start with a discussion of brane coupling renormalization. The
main complication in the higher-codimension case is the appearance
of power-law divergences, with all of the pitfalls and
complications which these entail for the low-energy description
\cite{usesandabuses}.

Consider the $(n+4)$-dimensional scalar field
\ba
    S = -\int \d^4 x\, \d^ny  \left[
    \frac12 (\partial_M \phi \, \partial^M \phi)
    + \frac 12 m_\ssB^2 \phi^2 +
    \delta^n(y) \, V_b(\phi) \right] \,,
\ea
with brane potential
\be
    V_b = -\frac 12 \lambda_2 \phi^2
    + \frac 14 \lambda_4 \phi^4 \,,
\ee
living in a flat space-time with metric
\ba
    \d s^2 = \eta_{\mu\nu}\d x^\mu \d x^\nu
    + \d r^2 +r^2 \gamma_{ab}(\theta) \d\theta^a \d\theta^b \,.
\ea
Here the $\theta^a$ are coordinates for the $n-1$ angular
directions, whose total volume we denote by $\varpi = \int
\d^{(n-1)} \theta \sqrt{\gamma}$. We focus for simplicity on
spherically symmetric solutions (independent of the angular
directions), although this assumption is not crucial (since higher
modes in the angular directions are regular at $r=0$).

As for codimension 2, the relation between the propagator, $G$, in
the presence of the brane coupling, and the propagator, $D$, in
its absence, is
\ba
    G_k(y;y') = D_k(y;y') + i \lambda_2
    \frac{D_k(y;0) D_k(0;y')}{1-i \lambda_2 D_k(0;0)} \,,
\ea
and as before the need for renormalization may be traced to the
divergence in $D_k(0;0)$. The nature of this divergence can be
divined from the mode sum giving the propagator, $D$, in the
absence of brane couplings
\ba
    \left[\Box - m_\ssB^2 \right] D(x,y;x',y')
    = i \delta^4(x-x') \delta^n(y-y') \,,
\ea
which, in brane-Fourier space,
\be
    D(x,y;x',y') = \int \frac{\d^4p}{(2\pi)^4} \;
    D_p(y;y') \, e^{ip\cdot(x-x')} \,,
\ee
has as solution
\ba
    D_p(x;x') = - i \int_0^{\infty} \frac{q^{n-1} \d q}{\varpi}
    \frac{1}{p^2 + m_\ssB^2 + q^2}
    \left[ \frac{1}{(q r)^{\nu}} J_{\nu}(q r) \right] \,
    \left[ \frac{1}{(q r')^{\nu}}J_{\nu}(q r') \right]
    + \cdots \,,
\ea
with $\nu=(n-2)/2$. The ellipses in this last equation represent
those terms involving the nontrivial angular modes.

Using the asymptotic form for $J_\nu$ in the limit $q r \ll 1$:
$J_{\nu}(qr) = (qr)^\nu/[\nu!2^{\nu}] + \mathcal{O}(qr)$, we find
\ba
    D_p(r=0;r'=0)= -  \frac{i}{\(\nu !\)^2 2^{2\nu}}
    \int_0^\infty \frac{q^{n-1} \d q}{\varpi}
    \frac{1}{m_\ssB^2+p^2+q^2}\,,
\ea
which diverges as a power of the UV cutoff, $\Lambda$, as
\ba
    D^{\tilde\Lambda}_p(0;0)&=& - \frac{i}{\(\nu !\)^2 2^{2\nu}}
    \int_0^{\tilde\Lambda} \frac{q^{n-1} \d
    q}{\varpi}\frac{1}{m_\ssB^2+p^2+q^2} \,.
    \\
    &=&-\frac{i}{\varpi \(\nu !\)^2
    2^{2\nu}} \left[ \frac{q^n}{n P^2}
    - \frac{q^{n+2}}{(n+2)P^4}
    \; {}_2F_1\(1,\frac{n+2}{2};
    \frac{n+4}{2},-\frac{q^2}{P^2}\)
    \right]_0^{\tilde\Lambda}\,,
\ea
where $P^2 = m_\ssB^2 + p^2$ and ${}_2F_1(a,b;c;z)$ denotes the
Hypergeometric function.

Our focus is on even $n$, $n = 2m$, in which case the
hypergeometric function can be simplified to the following
terminating series
\be
    {}_2F_1(1,m+1,m+2,z)
    = -(m+1) z^{-(m+1)} \left[ \log(1-z)
    +\sum_{j=1}^m \frac{z^j}{j} \right]\,.
\ee
Using this in the expression of the brane-brane propagator for
even codimensions, we get
\ba
    D^{\tilde\Lambda}_p(0;0) &=&
    \frac{i2^{2-m}}{\varpi [\Gamma(m)]^2}
    \Bigg[ \frac{(-)^{m}}{2} P^{2(m-1)}
    \log \left( 1+\frac{q^2}{P^2}\right)  \nn\\
    &&  \qquad\qquad\qquad\qquad
    +\frac12 \sum_{j=1}^{m-1} \frac{(-)^{j-m}}{j}
    \, q^{2j} P^{2(m-1-j)} \Bigg]_0^{\tilde \Lambda} \,.
\ea
For even codimension, $n = 2m$, we redefine $\Lambda^2 = \tilde
\Lambda^2 + P^2$, leading to
\ba
    D^\Lambda_p &=& - \frac{i2^{2-m}}{\varpi[\Gamma(m)]^2}
    (-P^2)^{m-1} \left[ \log \Lambda +
    \sum_{j=1}^{m-1} \frac{1}{2j} \left(
    1-\frac{\Lambda^2}{P^2} \right)^{j} \right]+ \text{(finite)}\,.
\ea
For odd codimensions, a similar argument gives
\ba
    D^\Lambda_p &=& - \frac{i2^{2-n}}{\varpi[\Gamma(n/2)]^2}
    \sum_{j=0}^{[n/2-1]} (-)^j \frac{P^{2j}
    \Lambda^{n-2-2j}}{n-2-2j}+ \text{(finite)}\,,
\ea
where $[n/2-1]$ denotes the largest integer smaller than $n/2-1$.

Renormalization proceeds as for codimension two, with the
requirement that
\ba
    \frac{\lambda_2(\Lambda)}{1-i
    \lambda_2(\Lambda) D^\Lambda_k(0,0)}
    = \frac{\lambda_2(\mu)}{1-i
    \lambda_2(\mu) D^\mu_k(0,0)}\,,
\ea
where $\mu$ is the renormalization scale, leading to the following
expression,
\ba \label{renormlambda2}
    \frac{1}{\lambda_2(\Lambda)} = \frac{1}{\bar
    \lambda_2(\mu)} + i \( D^\Lambda_k-D^\mu_k \) \,.
\ea

The divergence of propagator on the brane also induces divergences
in the expression of the 4-point function, which should be
absorbed by a renormalization of $\lambda_4$,
\ba
    G^{(4)}_{k_1,k_2,k_3,k_4}(y_1;y_2;y_3;y_4)
    &=& -6 i\,  \lambda_4 \left[
    \prod_{i=1}^4 G^{(2)}_{k_i}(y_i;0) \right]
    \delta^4\!\!\( \sum_i k_i\)\\
    &=&-6 i\,  \lambda_4 \left[ \prod_{i=1}^4
    \frac{D_{k_i}(y_i;0)}{1-i \lambda_2 D_{k_i}(0;0)} \right]
    \delta^4\!\!\( \sum_i k_i\)\,.
\ea
The quantity $\lambda_4/(1-i \lambda_2 D_{k_i}(0,0))^4$ is finite
if $\lambda_4$ is renormalized in the following way
\ba \label{renormlambda4}
    \lambda_4(\Lambda) = \frac{\bar \lambda_4}{
    \(1+i\bar \lambda_2\(D^\Lambda_k-D^\mu_k\)\)^4}\,.
\ea
Similar expressions can be found for higher-point couplings.

\subsection{Boundary condition and energy density}

We now turn to the classical solutions for $\phi(r)$, and the
boundary conditions which communicate the information of the brane
potential to the bulk theory. Just as in the main text the
singular form of the bulk solutions require us to regularize the
boundary condition by evaluating it at $r = \e$ rather than at
$r=0$. Smooth results are obtained as $\e \to 0$ once the bare
couplings are eliminated in terms of the renormalized couplings.

The classical solution to the bulk field equation that vanishes
far from the brane is
\ba
    \phi(r)= \bar \phi\, \frac{K_\nu (m_\ssB r)
    }{(m_\ssB r)^\nu} \,.
\ea
Integrating the equation of motion over the brane, we obtain the
boundary condition
\ba \label{bdycond_General}
    \varpi  \e^{n-1} \phi'_\e =
    -\lambda_2 \phi_\e+\lambda_4 \phi_\e^3 \,.
\ea
The energy density for such a field configuration is similarly
given by
\ba \label{energyDensityn}
    {\mathcal H}&=&v \int_\e^\infty r^{n-1} \d r \left[
    \frac12 (\partial_r \phi)^2 + \frac12 \, m_\ssB^2 \phi^2 \right]
    +U_b(\phi(\epsilon)) \nn\\
    &=&
    v \frac{m_\ssB^2}{2}\bar \phi^2\,
    \e^{n+1} \(m_\ssB \e\)^{-n}K_\nu(m_\ssB \e)K_{\nu+1}(m_\ssB \e)
    +U_b(\phi(\epsilon)) \,.
\ea
In general both of these last equations become finite once
expressed in terms of renormalized quantities, although the
cancellation becomes more regularization dependent in the
higher-codimension case due to the appearance there of power-law
divergences rather than logarithms. Rather than working this
through in complete generality, we restrict ourselves here to an
illustrative calculation for codimension three.

\subsection{Codimension-3}

For a codimension-3 brane the divergent part of the brane-brane
propagator goes as
\ba
    D^\Lambda_p = -\frac{2i\Lambda}{\pi \varpi}  \,,
\ea
and so the divergent part of the boundary condition
\eqref{bdycond_General} cancels identically if $2\Lambda/\pi =
\e$. The leading order part of the boundary condition becomes
\ba
    \bar \phi \(m_\ssB + \frac{\varpi}{\lt} -
    \mu + \frac{\pi \varpi^3\lf}{2\lt^4 m_\ssB^2}\, \bar \phi^2
    \)=0 \,,
\ea
where to simplify the notation we rescale $\mu \to \pi \mu/2$. The
system has solution $\bar\phi = 0$ as well as
\be
    \bar \phi^2 =
    - \left( \frac{2 \lt^4m_\ssB^2}{\pi \varpi^3 \lf}
    \right) m_{\rm{eff}}\,,
\ee
although the second solution is only possible when
\ba
    m_{\rm{eff}} = \(\frac{\varpi}{\lt}-\mu+m_\ssB\)<0\,.
\ea
These conclusions are consistent with the form of the energy
density, which in this case is
\ba
    \mathcal H= \left( \frac{\pi \varpi}{4m_\ssB^2} \right)
    \, m_{\rm{eff}} \, \bar \phi^2
    +\frac{\lf}{4}\(\frac{\varpi}{\lt m_\ssB}\)^4
    \(\frac{\pi}{2}\)^2 \bar
    \phi^4\,.
\ea
Notice that the criterion for having a nonzero v.e.v. in this case
depends more strongly on $m_\ssB$, relative to the codimension-2
case.

A similar argument can be made for higher codimensions. Notice
that for codimension-4 and higher, the propagator includes
sub-leading divergences which should also be renormalized. Doing
so, we recover a finite energy density with slightly different
criteria on having a nonzero v.e.v.

\section{Bulk Goldstone modes}
\label{appendixGoldModes}

A natural worry arises when the Higgs is regarded as a bulk scalar
while the Standard Model gauge bosons are confined to a brane.
Since the bulk $SU(2) \times U(1)$ rotations are not gauged, their
spontaneous breaking might be expected to bulk Goldstone modes,
corresponding to KK towers of bulk scalar modes whose lightest
members are massless (or with masses set by the KK scale, if the
global symmetries are broken by boundary conditions). Since only
three of these 4D KK states are eaten by the Higgs mechanism, the
remainder could survive and generate a potentially dangerous large
number of light states. In this section, we show that only three
massless Goldstone modes are produced, all of which are eaten by
the gauge fields on the brane.

We start with the argument in a nutshell: when choosing a specific
vacuum, such as the unitary gauge choice of the main text, one
expects Goldstone modes connecting to nearby vacua. Since all
vacua have the same profile in the extra dimensions, the Goldstone
modes also share this profile. The modes with the smallest energy
cost have only momentum along the brane directions, and so are
effectively already four-dimensional. These modes turn out to be
the self-localized states of those components of the Higgs doublet
that do not acquire a v.e.v.

To see this explicitly we repeat the calculation of the light
states in section~\ref{sec.SelfLocalizedState}, for the Higgs
doublet $H$. The equation of motion analogous to
Eq.~\eqref{SelfLocEq} is
\be
 \left[ \frac{1}{r^2}\left( (r\pd_r)^2 +\pd_\theta^2 \right)
 +\pd_\mu\pd^\mu - m_\ssB^2 \right]H
 = \frac{\delta_+(r)}{2\pi r}
 \left( -\lambda_2 + \lambda_4 H^\star H \right) H\,.
\ee
This equation may be linearized around the vacuum by setting
\be
 H = \left(\begin{array}{c}0\\ \varphi(r)  \end{array}\right)
 +\sum_{n\ge0}^\infty
 \left(\begin{array}{c}\zeta^n_1(r)+i\zeta^n_2(r)\\ \chi^n(r) + i\zeta_3^n(r) \end{array}
 \right)\sin (n \theta)\,,
\ee
with $\varphi(r)=\bar \phi K_0(m_\ssB r)$ and where we introduce
an infinite tower of excitation modes along the angular direction.
Each one of these modes satisfies the equations of motion
\ba \label{app:EOM1}
 \left[ \frac{1}{r} \pd_r \(r\pd_r\) -\frac{n^2}{r^2}
 -k^2 \right]\chi^n = \frac{\delta_+(r)}{2\pi r}
 \left( -\lambda_2 + 3 \lambda_4 \varphi^2(r)\right)\chi^n \\
 \label{app:EOM2}
 \left[ \frac{1}{r} \pd_r \(r\pd_r\) -\frac{n^2}{r^2}
 -k^2 \right]\zeta^n_i = \frac{\delta_+(r)}{2\pi r}
 \left( -\lambda_2 +\lambda_4
 \varphi^2(r)\right)\zeta^n_i\,,
\ea
where, as before, $k^2=m_\ssB^2-\omega^2$.

The field $\chi^0$ is the `physical' self-localized state,
discussed in the main text, and has a mass as calculated in
Eq.~\eqref{SelfLocalizedMass} with \eqref{l2eff}. The same goes
through for the zero mode of the other fields $\zeta_i^0$, but
taking into account the different factor of $\lambda_4$ between
equations \pref{app:EOM1} and \pref{app:EOM2}, their masses are
given by
\be \label{app:SelfLocalizedMass}
    \o^2_{\zeta^0} = m_\ssB^2 \left[ 1 -
    e^{-4\pi/\lambda_{\zeta^0}} \right] \,,
\ee
with now
\be \label{app:l2eff}
    \frac{1}{\lambda_{\zeta^0}} = \frac{1}{\lambda_{2\star}}
    \left[1+\frac{4\pi^2\bar\lambda_4\bar\phi^2}{\lambda_{2\star}^3}
    \right]\,.
\ee

In the broken phase, $\bar\phi$ is given by Eq.~\eqref{renvev}
which leads to
\be
 \frac{1}{\lambda_{\zeta^0}}=0\,,
\ee
showing there are three massless 4D Goldstone modes, $\zeta_i^0$.
The bulk profile of these modes is enforced by the boundary
condition imposed on the brane, and as argued in section
\ref{sec.TheModel}, choosing unitary gauge on the brane removes
these three massless states as they become `eaten' by the brane
gauge fields.

Turning now to the infinite tower of angular dependent modes
($n\ne0$), the profile of these modes is now of the form $\chi^n,
\zeta_i^n = N_i^n K_n(k r)$, where $N_i^n$ is the normalization
constant and we expect $k$ to be determined by the the boundary
condition \eqref{bdycond} which takes the form
\ba
 2\pi \left. r \partial_r
 \(\begin{array}{c} \chi^n(r) \\
 \zeta_i^n(r)\end{array}\)\right|_{\epsilon}
 =\(-\lambda_2 + \(\begin{array}{c} 3 \\
 1\end{array}\) \lambda_4 \varphi(r)^2\)
 \left. \(\begin{array}{c} \chi^n \\
 \zeta_i^n\end{array}\)\right|_{\epsilon}\,.
\ea
In the limit $\epsilon \to 0$, this reduces to
\ba
 2^n \pi N_i^n (k \epsilon)^{-n}\(n!+
 \frac{(n-1)!}{\log  (\epsilon \, m_\ssB
 e^\gamma/2) }\)=\mathcal{O}\((k \epsilon)^{-n+2}\)\,.
\ea
We see we must have $N_i^n =0$ if these modes are to remain
bounded, and so there are therefore no light modes of this form
having $\omega < m_\ssB$. All the remaining excitations along the
radial direction form a Kaluza-Klein tower of states starting at
the bulk mass $m_\ssB$ and are thus harmless. There are therefore
only three massless states $\zeta_i^0$ playing the role of
four-dimensional Goldstone modes, one self-localized massive mode
($\chi$) with mass $0<m<m_\ssB$ and a tower of Higgs excitations
with mass higher than the bulk mass.


\begin{thebibliography}{99}

\bibitem{SM}
For a recent summary of these issues see, C.P. Burgess and G.D.
Moore, {\it The Standard Model: A Primer}, Cambridge University
Press 2007.

\bibitem{ADD}
 N. Arkani-Hamed, S. Dimopoulos and G. Dvali, { Phys.\
Lett.} {\bf B429} (1998) 263 (hep-ph/9803315); Phys.\ Rev.\ {\bf
D59} (1999) 086004 (hep-ph/9807344);\\
I.~Antoniadis, N.~Arkani-Hamed, S.~Dimopoulos and G.~R.~Dvali,
  Phys.\ Lett.\  B {\bf 436}, 257 (1998)
  [arXiv:hep-ph/9804398].



\bibitem{RS}
 L. Randall, R. Sundrum, { Phys.\ Rev.\ Lett.} {\bf 83}
  (1999) 3370 [hep-ph/9905221], Phys.\ Rev.\ Lett.\ {\bf 83} (1999)
  4690 [hep-th/9906064].

\bibitem{Extra5DHiggs}
  A.~Pomarol and M.~Quiros,
  Phys.\ Lett.\  B {\bf 438} (1998) 255
  [arXiv:hep-ph/9806263];\\
  %
 A.~Delgado, A.~Pomarol and M.~Quiros,
  Phys.\ Rev.\  D {\bf 60} (1999) 095008
  [hep-ph/9812489];\\
  %
  I.~Antoniadis,
  Phys.\ Lett.\  B {\bf 246}, 377 (1990);\\
%
   T.~Gherghetta and A.~Pomarol,
  Nucl.\ Phys.\  B {\bf 586} (2000) 141
  [arXiv:hep-ph/0003129].

\bibitem{Extra6DHiggs}
L.~J.~Hall, Y.~Nomura and D.~R.~Smith,
  Nucl.\ Phys.\  B {\bf 639} (2002) 307
  [hep-ph/0107331];\\
  %
   I.~Gogoladze, Y.~Mimura and S.~Nandi,
  Phys.\ Lett.\  B {\bf 562} (2003) 307
  [hep-ph/0302176].

\bibitem{Gauge5DHiggs}
 G.~Burdman and Y.~Nomura,
  Nucl.\ Phys.\  B {\bf 656} (2003) 3
  [hep-ph/0210257];\\
%
    N.~Haba and Y.~Shimizu,
  Phys.\ Rev.\  D {\bf 67} (2003) 095001
  [Erratum-ibid.\  D {\bf 69} (2004) 059902]
  [hep-ph/0212166];\\
  %
   N.~Haba and T.~Yamashita,
  JHEP {\bf 0404} (2004) 016
  [hep-ph/0402157];\\
  %
   G.~Martinelli, M.~Salvatori, C.~A.~Scrucca and L.~Silvestrini,
  JHEP {\bf 0510} (2005) 037
  [hep-ph/0503179];\\
  %
   M.~Sakamoto and K.~Takenaga,
  Phys.\ Rev.\  D {\bf 75} (2007) 045015
  [hep-th/0609067].

\bibitem{Gauge6DHiggs}
 C.~A.~Scrucca, M.~Serone, L.~Silvestrini and A.~Wulzer,
  JHEP {\bf 0402} (2004) 049
  [hep-th/0312267];\\
%
  %
   C.~Biggio and M.~Quiros,
  Nucl.\ Phys.\  B {\bf 703} (2004) 199
  [hep-ph/0407348];\\
%
 Y.~Hosotani, S.~Noda and K.~Takenaga,
  Phys.\ Lett.\  B {\bf 607} (2005) 276
  [hep-ph/0410193];\\
%
  D.~Hernandez, S.~Rigolin and M.~Salvatori,
  arXiv:0712.1980 [hep-ph].



\bibitem{Dudas1}
  E.~Dudas, C.~Papineau and V.~A.~Rubakov,
  JHEP {\bf 0603}, 085 (2006)
  [arXiv:hep-th/0512276].

\bibitem{Dudas2}
  E.~Dudas and C.~Papineau,
  JHEP {\bf 0611}, 010 (2006)
  [arXiv:hep-th/0608054].





\bibitem{cod1case}
F.~Coradeschi, S.~De Curtis, D.~Dominici and J.~R.~Pelaez,
[arXiv:0712.0537 [hep-th]].

\bibitem{JJ}
 J.~Vinet and J.~M.~Cline,
  Phys.\ Rev.\  D {\bf 70} (2004) 083514
  [hep-th/0406141].

\bibitem{UVCaps}
 M.~Peloso, L.~Sorbo and G.~Tasinato,
  Phys.\ Rev.\  D {\bf 73} (2006) 104025
  [hep-th/0603026];\\
  %
    E.~Papantonopoulos, A.~Papazoglou and V.~Zamarias,
  JHEP {\bf 0703} (2007) 002
  [hep-th/0611311];\\
  %
  B.~Himmetoglu and M.~Peloso,
  Nucl.\ Phys.\  B {\bf 773} (2007) 84
  [hep-th/0612140];\\
  %
   N.~Kaloper and D.~Kiley,
  JHEP {\bf 0705} (2007) 045
  [hep-th/0703190];\\
  %
 C.~P.~Burgess, D.~Hoover and G.~Tasinato,
  JHEP {\bf 0709} (2007) 124
  [arXiv:0705.3212 [hep-th]];\\
  %
  M.~Minamitsuji and D.~Langlois,
  Phys.\ Rev.\  D {\bf 76} (2007) 084031
  [arXiv:0707.1426 [hep-th]];\\
  %
   F.~Arroja, T.~Kobayashi, K.~Koyama and T.~Shiromizu,
  JCAP {\bf 0712} (2007) 006
  [arXiv:0710.2539 [hep-th]];\\
%
C.~Bogdanos, A.~Kehagias and K.~Tamvakis,
  Phys.\ Lett.\  B {\bf 656} (2007) 112
  [arXiv:0709.0873 [hep-th]].

\bibitem{cosmicstrings}
A. Vilenkin, Phys. Rev. {\bf D23} (1981) 852;\\
%
R.~Gregory and C.~Santos,
Phys.\ Rev.\ D {\bf 56}, 1194 (1997) [gr-qc/9701014].

\bibitem{conicaldefects}
S.~M.~Carroll and M.~M.~Guica,
[hep-th/0302067];\\

\bibitem{SLED}
 Y.~Aghababaie, C.~P.~Burgess, S.~L.~Parameswaran and F.~Quevedo,
  Nucl.\ Phys.\  B {\bf 680} (2004) 389
  [hep-th/0304256].

\bibitem{nonconicalcod2}
 G.~W.~Gibbons, R.~Guven and C.~N.~Pope,
  Phys.\ Lett.\  B {\bf 595} (2004) 498
  [hep-th/0307238];
  %
 Y.~Aghababaie {\it et al.},
  JHEP {\bf 0309} (2003) 037
  [hep-th/0308064];\\
  %
    P.~Bostock, R.~Gregory, I.~Navarro and J.~Santiago,
  Phys.\ Rev.\ Lett.\  {\bf 92} (2004) 221601
  [hep-th/0311074];\\
  %
   C.~P.~Burgess, F.~Quevedo, G.~Tasinato and I.~Zavala,
  JHEP {\bf 0411} (2004) 069
  [hep-th/0408109];
  %
  J.~Vinet and J.~M.~Cline,
  Phys.\ Rev.\  D {\bf 71} (2005) 064011
  [hep-th/0501098].\\

\bibitem{GW}
  W.~D.~Goldberger and M.~B.~Wise,
  Phys.\ Rev.\  D {\bf 65} (2002) 025011
  [arXiv:hep-th/0104170].

\bibitem{CdR}
  C.~de Rham,
  JHEP {\bf 0801} (2008) 060
  [arXiv:0707.0884 [hep-th]];\\
  C.~de Rham,
  AIP Conf.\ Proc.\  {\bf 957}, 309 (2007)
  [arXiv:0710.4598 [hep-th]].



\bibitem{warpedlocalscalar}
D.~Langlois and M.~Sasaki,
  Phys.\ Rev.\  D {\bf 68} (2003) 064012
  [hep-th/0302069].

\bibitem{dimtrans}
 W.~Frank, D.~J.~Land and R.~M.~Spector,
  Rev.\ Mod.\ Phys.\  {\bf 43} (1971) 36;\\
  %
   G.~Parisi and F.~Zirilli,
  J.\ Math.\ Phys.\  {\bf 14} (1973) 243;\\
  %
H.~E.~Camblong, L.~N.~Epele, H.~Fanchiotti and C.~A.~Garcia Canal,
  Phys.\ Rev.\ Lett.\  {\bf 85} (2000) 1590
  [hep-th/0003014];\\
  %
 S.~R.~Beane, P.~F.~Bedaque, L.~Childress, A.~Kryjevski, J.~McGuire and U.~v.~Kolck,
  Phys.\ Rev.\  A {\bf 64} (2001) 042103
  [quant-ph/0010073].

\bibitem{braaten}
E. Braaten and D. Phillips, [hep-th/0403168].

\bibitem{GB}
See for instance,
  C.~P.~Burgess,
  Phys.\ Rept.\  {\bf 330} (2000) 193
  [arXiv:hep-th/9808176].

\bibitem{5Dbranegaugebulkhiggs}
For a discussion of these issues in the codimension-1 case, see
A.~Kehagias and K.~Tamvakis,
  Phys.\ Lett.\  B {\bf 628} (2005) 262
  [hep-th/0507130].

\bibitem{GIM}
S.~L.~Glashow, J.~Iliopoulos and L.~Maiani,
  Phys.\ Rev.\  D {\bf 2} (1970) 1285.

\bibitem{oblique}
M.~E.~Peskin and T.~Takeuchi,
  Phys.\ Rev.\ Lett.\  {\bf 65} (1990) 964;
  %
  Phys.\ Rev.\  D {\bf 46} (1992) 381;\\
%
  G.~Altarelli and R.~Barbieri,
  Phys.\ Lett.\  B {\bf 253} (1991) 161;\\
  %
  G.~Altarelli, R.~Barbieri and S.~Jadach,
  Nucl.\ Phys.\  B {\bf 369} (1992) 3
  [Erratum-ibid.\  B {\bf 376} (1992) 444];\\
  %
  C.~P.~Burgess, S.~Godfrey, H.~Konig, D.~London and I.~Maksymyk,
  Phys.\ Lett.\  B {\bf 326}, 276 (1994)
  [hep-ph/9307337];
  %
  Phys.\ Rev.\  D {\bf 49}, 6115 (1994)
  [hep-ph/9312291].

\bibitem{custodial}
S. Weinberg, Phys.\ Rev.\ D{\bf 19}, 1277 (1979);\\
%
L. Susskind, Phys.\ Rev.\ D{\bf 20}, 2619 (1979);\\
%
P. Sikivie, L. Susskind, M. Voloshin and V. Sakharov, Nucl.\
Phys.\ B{\bf 173} (1980) 189;\\
%
H.~Georgi and D.~B.~Kaplan,
  Phys.\ Lett.\  B {\bf 145} (1984) 216.

\bibitem{unitaritybound}
J.M. Cornwall, D.N. Levin and G. Tiktopoulos,
 Phys.\ Rev.\ D{\bf 10} 1145 (1974);\\
 %
 B.W. Lee, C. Quigg and H. Thacker,
 Phys.\ Rev.\ D{\bf 16} 1519 (1977);\\
 %
 M. Veltman, Acta.\ Phys.\ Pol.\ B{\bf 8} 475 (1977).

\bibitem{usesandabuses}
C.~P.~Burgess and D.~London,
  Phys.\ Rev.\  D {\bf 48} (1993) 4337
  [hep-ph/9203216];
%
  Phys.\ Rev.\ Lett.\  {\bf 69} (1992) 3428.

\bibitem{MSLED}
C.~P.~Burgess, J.~Matias and F.~Quevedo,
  Nucl.\ Phys.\  B {\bf 706} (2005) 71
  [hep-ph/0404135].

\bibitem{ADDphen}
G.~F.~Giudice, R.~Rattazzi and J.~D.~Wells,
  Nucl.\ Phys.\  B {\bf 544} (1999) 3
  [hep-ph/9811291];\\
  %
  T.~Han, J.~D.~Lykken and R.~J.~Zhang,
  Phys.\ Rev.\  D {\bf 59} (1999) 105006
  [hep-ph/9811350].

\bibitem{SLEDphen}
D.~Atwood, C.~P.~Burgess, E.~Filotas, F.~Leblond, D.~London and
I.~Maksymyk,
  Phys.\ Rev.\  D {\bf 63} (2001) 025007
  [hep-ph/0007178];\\
  %
  I.~Antoniadis and K.~Benakli,
  Int.\ J.\ Mod.\ Phys.\  A {\bf 15} (2000) 4237
  [hep-ph/0007226];\\
  %
  J.~L.~Hewett and D.~Sadri,
  Phys.\ Rev.\  D {\bf 69} (2004) 015001
  [hep-ph/0204063];\\
%
    P.~H.~Beauchemin, G.~Azuelos and C.~P.~Burgess,
  J.\ Phys.\ G {\bf 30} (2004) N17
  [hep-ph/0407196];
  %
  J.\ Phys.\ G {\bf 31} (2005) 1
  [hep-ph/0401125].

\end{thebibliography}
\end{document}